\documentclass[journal, letterpaper,draftcls,onecolumn,12pt,twoside]{IEEEtran}
\usepackage{cite,graphicx,amsmath,amssymb,array,stfloats,cuted,midfloat,psfrag}
\newtheorem{theorem}{\textbf{Theorem}}
\newtheorem{definition}{\textbf{Definition}}
\newtheorem{lemma}{\textbf{Lemma}}

\newtheorem{Corollary}{\textbf{Corollary}}

\begin{document}
\title{Robust Distributed Source Coding}
\author{Jun Chen~\IEEEmembership{Member,~IEEE}, Toby
Berger,~\IEEEmembership{Fellow,~IEEE}
\thanks{
J. Chen and T. Berger are with the School of Electrical and
Computer Engineering, Cornell University, Ithaca, NY 14853 USA
(email: jc353@cornell.edu, berger@ece.cornell.edu). This work was
supported in part by NSF Grant CCR-033 0059 and a grant from the
National Academies Keck Futures Initiative (NAKFI).}}
\markboth{SUBMITTED TO IEEE TRANSACTIONS ON INFORMATION THEORY}
{CHEN \textit{et al.}:ROBUST DISTRIBUTED SOURCE CODING} \maketitle
\date{}

\begin{abstract}
We consider a distributed source coding system in which several
observations are communicated to the decoder using limited
transmission rate. The observations must be separately coded. We
introduce a robust distributed coding scheme which flexibly trades
off between system robustness and compression efficiency. The
optimality of this coding scheme is proved for various special
cases.

\end{abstract}
\section{Introduction}
There are many situations in which data collected at several sites
must be transmitted to a common point for subsequent processing.
Via clever encoding techniques, it is possible to capitalize on
the correlation between data received at the various sites even
though each encoder operates with no or only partial knowledge of
the data received at the other sites. Slepian and Wolf \cite{SW}
proved a coding theorem for two correlated memoryless sources with
separate encoders. They dealt with the case where the decoder must
reproduce two source outputs with arbitrary small error
probability. Their results were extended to arbitrary number of
discrete sources with ergodic memory and countably infinite
alphabets by Cover \cite{Cover}. Based on the results of Slepian
and Wolf, Wyner and Ziv \cite{Wyner-Ziv} extended rate-distortion
theory to the case in which side information is present at the
decoder. Berger \cite{Berger} and Tung \cite{Tung} generalized the
Slepian-Wolf problem by considering general distortion criteria on
the source reconstruction. The complete characterization of the
rate-distortion region is unknown except for the special case
where one of two source outputs must be reconstructed with an
arbitrary small error probability and the other must have an
average distortion smaller than a prescribed level
\cite{BergerYeung1}. Oohama \cite{OohamaGaussian} studied the
rate-distortion region for correlated memoryless Gaussian sources
and squared distortion measures. He demonstrated that the inner
bound of the rate-distortion region obtained by Berger and Tung is
partially tight in the Gaussian case. Viswanath \cite{Viswanath}
characterized the sum-rate distortion function of Gaussian
multiterminal source coding problem for a class of quadratic
distortion metrics. A closely related problem, called the remote
source coding problem or the CEO problem, has been studied in
\cite{GP,FG,BergerCEO,GaussianCEO,Draper}. Oohama \cite{OohamaCEO}
derived the sum-rate distortion function for the quadratic
Gaussian CEO problem when there are infinite encoders and the SNRs
at all the encoders are identical. It was observed by Chen
\textit{et al.} \cite{ChenCEO} that Oohama's converse yields a
tight upper bound on the sum-rate distortion function even when
the number of encoders are finite. They also computed the
achievable region for the general quadratic Gaussian CEO problem.
Recently, Oohama \cite{OohamaU} and Prabhakaran \textit{et al.}
\cite{TseCEO} showed that this achievable region is indeed the
rate-distortion region.

Another important class of source coding problems is called
multiple description problem. In the multiple description problem,
the total available bit rate is split between (say) two channels
and either channel may be subject to failure. It is desired to
allocate rate and coded representations between the two channels,
such that if one channel fails, an adequate reconstruction of the
source is possible, but if both channels are available, an
improved reconstruction over the single-channel reception results.
This problem was posed by Gersho, Witsenhausen, Wolf, Wyner, Ziv
and Ozarow in 1979. Early contributions to this problem can be
found in Witsenhausen \cite{Witsenhausen}, Wolf, Wyner and Ziv
\cite{WWZ}, Ozarow \cite{Ozarow} and Witsenhausen and Wyner
\cite{WW}. The first general result was El Gamal and Cover's
achievable region for two channels \cite{GamalCover}. Ahlswede
\cite{Ahlswede} showed that in the ``no excess rate" case, El
Gamal and Cover's region is tight. Zhang and Berger
\cite{ZhangBerger} exhibit a simple counterexample that shows El
Gamal and Cover's region is not always tight in the case of an
excess rate. Further results can be found in
\cite{ref23,ref27,ref29,ref30,ref33,ref34,ref35,ref42}.

Distributed source coding problems of the Slepian-Wolf type and
its extensions emphasize the compression efficiency of coding
system but ignore the system robustness. A distributed source
coding scheme which is optimal in the sense of compression
efficiency can be very sensitive to the encoder failure, i.e., the
performance of the whole system may degenerate dramatically when
one of the encoders is subject to a failure. On the other hand,
multiple description problem does consider the system robustness.
But it is essentially a centralized source coding problem whose
coding schemes in general can not be applied in the distributed
source coding scenario. So it is of interest to study robust
distributed source coding scheme, which is able to trade off
between two important parameters: system robustness and
compression efficiency. 

\section{System Model and Problem Formulation}
Consider the distributed source coding system shown in Fig. 1. Let
$\{X(t), Y_1(t), Y_2(t)\}_{t=1}^{\infty}$ be temporally memoryless
source with instantaneous joint probability distribution $P(x,
y_1, y_2)$ on $\mathcal{X}\times \mathcal{Y}_1\times
\mathcal{Y}_2$, where $\mathcal{X}$ is the common alphabet of the
random variables $X(t)$ for $t=1, 2, \cdots$, $\mathcal{Y}_i\
(i=1, 2)$ is the common alphabet of the random variables $Y_i(t)$
for $t=1, 2, \cdots$. $\{X(t)\}_{t=1}^{\infty}$ is the target data
sequence which can not be observed directly. Instead, two
corrupted versions of $\{X(t)\}_{t=1}^{\infty}$, i.e.,
$\{Y_1(t)\}_{t=1}^{\infty}$ and $\{Y_2(t)\}_{t=1}^{\infty}$, are
observed by encoder 1 and encoder 2 respectively. Encoder $i$
encodes a block $y_i^n=[y_i(1), \cdots, y_i(n)]$ of length $n$
from its observed data using a source code
$c_i^{(n)}=f_{E,i}^{(n)}(y_i^n)$ of rate
$\frac{1}{n}\log|\mathcal{C}_i^{(n)}|$. Decoder $i$ reconstructs
the target sequence $x^n=[x(1), \cdots, x(n)]$ by implementing a
mapping $f^{(n)}_{D, i}:\ \mathcal{C}^{(n)}_{i}\rightarrow
\mathcal{X}^n, \quad i=1,2$. Decoder 3 reconstructs the target
sequence $x^n=[x(1), \cdots, x(n)]$ by implementing a mapping
$f^{(n)}_{D, 3}:\ \mathcal{C}_{1}^{(n)}\times
\mathcal{C}_{2}^{(n)}\rightarrow \mathcal{X}^n$.

\begin{figure} \label{robustmodel}
\centering
\begin{psfrags}
\psfrag{x}[c]{$X$}%
\psfrag{y1}[c]{$Y_1$}%
\psfrag{y2}[c]{$Y_2$}%
\psfrag{en1}[c]{Encoder $1$}%
\psfrag{en2}[c]{Encoder $2$}%
\psfrag{de1}[c]{Decoder $1$}%
\psfrag{de2}[c]{Decoder $2$}%
\psfrag{de3}[c]{Decoder $3$}%
\psfrag{xhat1}[l]{$\hat X_1\sim D_1$}%
\psfrag{xhat2}[l]{$\hat X_2\sim D_2$}%
\psfrag{xhat3}[l]{$\hat X_3\sim D_3$}%
\psfrag{r1}[c]{$R_1$}%
\psfrag{r2}[c]{$R_2$}%
\includegraphics[scale=0.7]{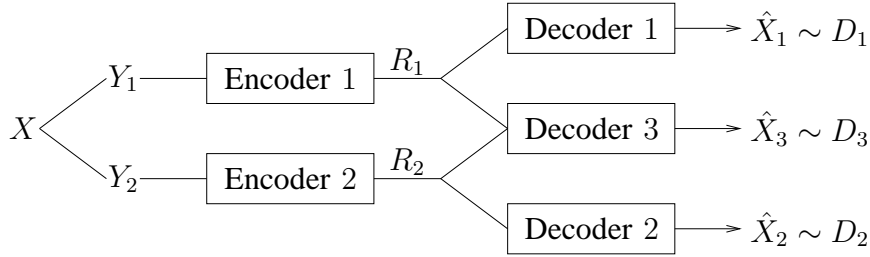}
\end{psfrags}
\caption{Model of robust distributed source coding system}
\end{figure}

\begin{definition}
The quintuple $(R_1, R_2, D_1, D_2, D_3)$ is called achievable, if
for any $\epsilon>0$, there exists an $n_0$ such that for all
$n>n_0$ there exist encoders:
\begin{equation*}
  f_{E, i}^{(n)}:
  \mathcal{Y}_i^{n}\rightarrow\mathcal{C}_i^{(n)} \quad
  \log|\mathcal{C}_i^{(n)}|\leq n(R_i+\epsilon) \quad i=1,2
\end{equation*}
and decoders:
\begin{equation*}
f^{(n)}_{D, i}:\ \mathcal{C}_{i}^{(n)}\rightarrow \mathcal{X}^n
\quad i=1,2
\end{equation*}
\begin{equation*}
  f^{(n)}_{D, 3}:\ \mathcal{C}_{1}^{(n)}\times
\mathcal{C}_{2}^{(n)}\rightarrow \mathcal{X}^n
\end{equation*}
such that for $\hat
X^n_i=f^{(n)}_{D,i}\left(f^{(n)}_{E,i}(Y^n_i)\right), i=1,2$, and
for $\hat
X^n_3=f^{(n)}_{D,3}\left(f^{(n)}_{E,1}(Y^n_1),f^{(n)}_{E,2}(Y^n_2)\right)$,
\begin{equation*}
\frac{1}{n}\mathbb{E}\sum\limits_{t=1}^n d(X(t), \hat
X_i(t))<D_i+\epsilon \quad i=1,2,3.
\end{equation*}
Here
$d(\cdot,\cdot):\mathcal{X}\times\mathcal{X}\rightarrow[0,d_{\max}]$
is a given distortion measure.

Let $\mathcal{Q}$ denote the set of all achievable quintuples.
\end{definition}

Remark:
\begin{enumerate}
  \item Our model applies to many different scenarios such as
the nonergodic link failures from some encoders to the decoder or
the malfunction of some encoders;
  \item We restrict
our treatment to the case of two encoders just for simplifying the
notations. Most of our results can be extended in a
straightforward way to the case of arbitrary number of encoders;
\end{enumerate}

Our model was first introduced by Ishwar \textit{et al.} in
\cite{Ishwar}. An analogous problem called multilevel diversity
coding has been studied in
\cite{Roche,Yeung,RocheYeung,YeungZhang}. But it is a centralized
source coding problem since all the encoders have the same
observation. A distributed version of multilevel diversity coding
was introduced in \cite{YeungZhang2}, where only the case of
lossless source coding was treated.

The rest of this paper is divided into four sections. Section III
uses some examples to motivate the results. In Section IV, we
first consider two different scenarios, namely, the centralized
source coding and the distributed source coding, for which the
corresponding coding schemes are established. Then we propose a
unified approach by developing a coding scheme based on the idea
of common information. In Section V, the case of correlated
memoryless Gaussian observations and squared distortion measure is
studied in detail. The inner bound and outer bound of the rate
distortion region are established. We show that in various special
cases the complete characterization of the rate distortion region
is possible. Finally Section VI concludes the paper.

\section{Motivations and Examples}

Let $D_{\max}=\min_{x_0\in\mathcal{X}}\mathbb{E}d(X,x_0)$. Our
problem reduces to the CEO problem if $\min(D_1,D_2)\geq D_{\max}$
and reduces the multiple description problem if there exist
deterministic functions $f_i$ $(i=1, 2)$ such that
$X(t)=f_i(Y_i(t))$ with probability one for $t=1,2,\cdots$. So it
is instructive to review the coding schemes for the CEO problem
and multiple description problem.

For the CEO problem, the fidelity criterion is only imposed on the
reconstruction of the target sequence at decoder 3. The largest
known achievable rate distortion region for the CEO problem is the
set\footnote{By a timesharing argument, the convex hull of this
region is also achievable.} of $(R_1,R_2,D_3)$ for which there
exist random variables $W_1, W_2$ jointly distributed with the
generic source variables $X, Y_1$ and $Y_2$ such that
\begin{itemize}
  \item [(i)] $W_1\rightarrow Y_1\rightarrow (X, Y_2, W_2)$\footnote{$A\rightarrow B\rightarrow C$ means $A,B,$ and $C$ form a Markov chain, i.e., $A$
  and $C$ are independent conditioned on $B$.} and $W_2\rightarrow Y_2\rightarrow (X, Y_1, W_1)$.
  \item [(ii)] $R_1\geq I(Y_1;W_1|W_2), R_2\geq I(Y_2;W_2|W_1), R_1+R_2\geq I(Y_1,Y_2;W_1,W_2)$.
  \item [(iii)] There exist a function $f:\mathcal{W}_1\times\mathcal{W}_2\rightarrow\mathcal{X}$
such that $\mathbb{E} d(X, {\hat X})\leq D$,  where ${\hat
X}=f(W_1, W_2)$.
\end{itemize}

The proof of the achievability of this rate distortion region is
based on the idea  of random binning. The main feature of the
random binning coding scheme is outlined as follows:

\textit{There are many bins at each encoder and many codewords in
each bin. Instead of directly sending the codeword, each encoder
sends the index of bin which contains the codeword that this
encoder wants to reveal to the decoder. Upon receiving the indices
of bins from all the encoders, the decoder picks one codeword from
each bin such that these codewords are jointly typical. }

There are two important parameters for each encoder: the number of
bins and the number of codewords. Roughly speaking, the number of
bins determines the rate of the encoder while the number of
codewords is associated with the description ability of the
encoder. When the system is optimized in the sense of compression
efficiency, the number of bins is minimized at each encoder if the
number of its codewords is fixed (or equivalently, the number of
codewords is maximized at each encoder if the number of its bins
is fixed). Note: there exists a tradeoff between the maximum
number of codewords at different encoders if the number of bins is
fixed at each encoder (or equivalently,  a tradeoff between the
minimum number of bins at different encoders if the number of
codewords is fixed at each encoder). But intuitively this
optimization is achieved at the price of sacrificing the
robustness of the whole system: if the decoder only receives the
data from one of the encoders, then it may not be able to recover
the correct codeword since the decoder only gets a bin index from
one encoder and there are many codewords in that bin. Clearly, if
there is only one codeword in each bin, then the decoder is able
to recover the codeword as long as the bin index is received.
Actually now the encoding scheme reduces to the conventional lossy
source encoding and the joint decoding scheme becomes the separate
decoding. In general, we can improve the robustness of the
distributed source coding system by reducing the number of
codewords in each bin, which is a way to trade the compression
efficiency for the system robustness. This is essentially the main
idea of the robust distributed source coding scheme proposed by
Ishwar, Puri, Pradhan and Ramchandran \cite{Ishwar}, which we will
refer to as the IPPR scheme. The achievable rate distortion region
of IPPR scheme for our model is the set of $(R_1,R_2,D_1,D_2,D_3)$
for which there exist random variables $W_1, W_2$ jointly
distributed with the generic source variables $X, Y_1$ and $Y_2$
such that
\begin{itemize}
  \item [(i)] $W_1\rightarrow Y_1\rightarrow (X, Y_2, W_2)$ and $W_2\rightarrow Y_2\rightarrow (X, Y_1, W_1)$.
  \item [(ii)] $R_1\geq I(Y_1;W_1), R_2\geq I(Y_2;W_2)$.
  \item [(iii)] There exist functions $f_i:\mathcal{W}_i\rightarrow\mathcal{X} \quad
  (i=1,2)$, and
  $f_3:\mathcal{W}_1\times\mathcal{W}_2\rightarrow\mathcal{X}$
such that $\mathbb{E} d(X, {\hat X}_i)\leq D_i \quad (i=1,2,3)$,
where ${\hat X}_1=f_1(W_1)$, ${\hat X}_2=f_2(W_2)$ and ${\hat
X}_3=f_3(W_1, W_2)$.
\end{itemize}

We need the following definition before discussing the properties
of the IPPR scheme.
\begin{definition}
\begin{eqnarray*}
&&D^*_1(R_1,R_2)=\min\{D_1:(R_1,R_2,D_1,D_{\max},D_{\max})\in\mathcal{Q}\},\\
&&D^*_2(R_1,R_2)=\min\{D_2:(R_1,R_2,D_{\max},D_2,D_{\max})\in\mathcal{Q}\},\\
&&D^*_3(R_1,R_2)=\min\{D_3:(R_1,R_2,D_{\max},D_{\max},D_3)\in\mathcal{Q}\}.
\end{eqnarray*}
It is clear that $D^*_1(R_1,R_2)$ does not depend on $R_2$ and
$D^*_2(R_1,R_2)$ does not depend on $R_1$, so we shall denote them
by $D^*_1(R_1)$ and $D^*_2(R_2)$ respectively. $D^*_1(R_1)$ and
$D^*_2(R_2)$ are essentially the distortion-rate functions with
noisy observations \cite{ZivWolf}\cite{BergerText}, i.e.,
\begin{eqnarray*}
D^*_i(R_i)=\min\limits_{\substack{\hat
X_i\in\mathcal{X}:X\rightarrow Y_i\rightarrow \hat X_i\\I(Y_i;\hat
X_i)\leq R_i}}\mathbb{E}d(X,\hat X_i),\quad i=1,2.
\end{eqnarray*}
\end{definition}

The IPPR scheme is of special interest in the sense that given
rate tuple $(R_1,R_2)$, it can achieve $D^*_1(R_1)$ and
$D^*_2(R_2)$ at decoder 1 and decoder 2 respectively as shown by
the following argument:

Let $W^*_1(R_1), W^*_2(R_2)\in\mathcal{X}$ be the random variables
jointly distributed with $X, Y_1$ and $Y_2$ such that
$W^*_1(R_1)\rightarrow Y_1\rightarrow (X, Y_2, W^*_2(R_2))$ and
$W^*_2(R_2)\rightarrow Y_2\rightarrow (X, Y_1, W^*_1(R_1))$ with
$I(Y_i;W^*_i(R_i))\leq R_i$ and
$\mathbb{E}d(X,W^*_i(R_i))=D^*_i(R_i), i=1,2$.

By the IPPR scheme,
$(R_1,R_2,D^*_1(R_1),D^*_2(R_2),\min_{g:\mathcal{X}\times\mathcal{X}\rightarrow\mathcal{X}}
Ed(X,g(W^*_1(R_1),W^*_2(R_2))))$ is achievable, where
$\min_{g:\mathcal{X}\times\mathcal{X}\rightarrow\mathcal{X}}
\mathbb{E}d(X,g(W^*_1(R_1),W^*_2(R_2)))\leq\min(D^*_1(R_1),D^*_2(R_2))$
and the inequality is strict for most cases of interest.

Now we shall study the quadratic Gaussian case to get a concrete
feeling about the IPPR scheme. Suppose $X\sim
\mathcal{N}(0,\sigma^2_X)$, $Y_i=X+N_i$, $N_i\sim
\mathcal{N}(0,\sigma^2_{N_i})$, $i=1,2$. Here $X, N_1$ and $N_2$
are all independent. Let $W_1=Y_1+T_1, W_2=Y_2+T_2$, where
$T_i\sim \mathcal{N}(0,\sigma^2_{T_i})$, $i=1,2,$ are independent
of $X, N_1$ and $N_2$.

\begin{figure}[htb] \label{modeldssL}
\centering {\includegraphics[scale=0.5]{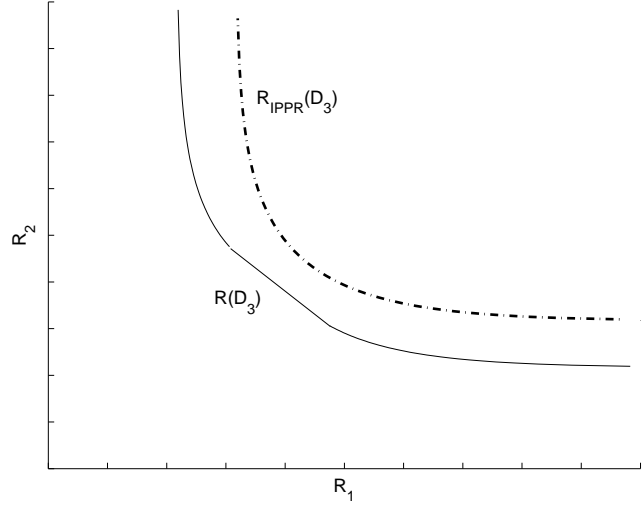}} \caption{IPPR
scheme}
\end{figure}

By the IPPR scheme, for any
$(R_1,R_2)\in\mathcal{R}_{IPPR}(D_1,D_2,D_3)$, we have
$(R_1,R_2,D_1,D_2,D_3)\in\mathcal{Q}$, where
\begin{equation*}
  \mathcal{R}_{IPPR}(D_1,D_2,D_3)=\bigcup\limits_{(\sigma_{T_1}^2,\sigma_{T_2}^2)\in\Sigma'(D_1,D_2,D_3)}\mathcal{R}'(\sigma_{T_1}^2,\sigma_{T_2}^2),
\end{equation*}
\begin{eqnarray*}
\mathcal{R}'(\sigma_{T_1}^2,\sigma_{T_2}^2)=\left\{(R_1,R_2):
R_1\geq
I(Y_1;W_1)=\frac{1}{2}\log\frac{\sigma_X^2+\sigma_{N_1}^2+\sigma_{T_1}^2}{\sigma_{T_1}^2},\right.\\
\left. R_2\geq
I(Y_2;W_2)=\frac{1}{2}\log\frac{\sigma_X^2+\sigma_{N_2}^2+\sigma_{T_2}^2}{\sigma_{T_2}^2}\right\}
\end{eqnarray*}
and
\begin{eqnarray*}
\Sigma'(D_1,D_2,D_3)=\left\{(\sigma_{T_1}^2,\sigma_{T_2}^2):
D_1\geq \mathbb{E}(X-\mathbb{E}(X|W_1))^2=
\left(\frac{1}{\sigma^2_X}+\frac{1}{\sigma_{N_1}^2+\sigma_{T_1}^2}\right)^{-1},\right.\\D_2\geq
\mathbb{E}(X-\mathbb{E}(X|W_2))^2=
\left(\frac{1}{\sigma^2_X}+\frac{1}{\sigma_{N_2}^2+\sigma_{T_2}^2}\right)^{-1},\\
\left.D_3\geq \mathbb{E}(X-\mathbb{E}(X|W_1,W_2))^2=
\left(\frac{1}{\sigma^2_X}+\frac{1}{\sigma_{N_1}^2+\sigma_{T_1}^2}+\frac{1}{\sigma_{N_2}^2+\sigma_{T_2}^2}\right)^{-1}\right\}.
\end{eqnarray*}
It was computed in \cite{ChenCEO} that
\begin{equation*}
R(D,\sigma^2_N)=\frac{1}{2}\log\frac{\sigma_X^4}{D\sigma_X^2-\sigma_X^2\sigma_{N}^2+D\sigma_{N}^2},
\end{equation*}
where $R(D,\sigma^2_N)$ the sum-rate distortion function of the
one-encoder quadratic Gaussian CEO problem\footnote{The
one-encoder CEO problem is the same as the problem of lossy source
coding with noisy observations}. So we have
\begin{equation*}
D^*_i(R_i)=\frac{\sigma^4_X\exp(-2R_i)+\sigma^2_X\sigma^2_{N_i}}{\sigma^2_X+\sigma^2_{N_i}},\quad
i=1,2.
\end{equation*}
It is easy to check that $\left(R_1,R_2\right)\in
\mathcal{R}_{IPPR}(D^*_1(R_1)$,$D^*_2(R_2),D_3)$ given
$D_3\geq(1/{D^*_1(R_1)}+1/{D^*_2(R_2)}-1/{\sigma_X^2})^{-1}$.
Hence IPPR scheme can indeed achieve $D^*_1(R_1)$ and $D^*_2(R_2)$
at decoder 1 and decoder 2 respectively in the quadratic Gaussian
case.

Let
$\mathcal{R}(D_3)=\left\{(R_1,R_2):(R_1,R_2,\sigma^2_X,\sigma^2_X,D_3)\in\mathcal{Q}\right\}$.
It was shown in \cite{OohamaU,TseCEO} that
\begin{eqnarray*}
\mathcal{R}(D_3)=\bigcup\limits_{(\sigma_{T_1}^2,\sigma_{T_2}^2)\in\Sigma''(D_3)}\mathcal{R}''(\sigma_{T_1}^2,\sigma_{T_2}^2),
\end{eqnarray*}
where
\begin{eqnarray*}
\mathcal{R}''(\sigma_{T_1}^2,\sigma_{T_2}^2)=\left\{(R_1,R_2):
R_1\geq
I(Y_1;W_1|W_2)=\frac{1}{2}\log\frac{(\sigma_X^2+\sigma_{N_1}^2+\sigma_{T_1}^2)
(\sigma_X^2+\sigma_{N_2}^2+\sigma_{T_2}^2)-\sigma_X^4}{\sigma_X^2\sigma_{T_1}^2+\sigma_{N_2}^2\sigma_{T_1}^2+\sigma_{T_1}^2\sigma_{T_2}^2}\right.,\\
R_2\geq
I(Y_2;W_2|W_1)=\frac{1}{2}\log\frac{(\sigma_X^2+\sigma_{N_1}^2+\sigma_{T_1}^2)
(\sigma_X^2+\sigma_{N_2}^2+\sigma_{T_2}^2)-\sigma_X^4}{\sigma_X^2\sigma_{T_2}^2+\sigma_{N_1}^2\sigma_{T_2}^2+\sigma_{T_1}^2\sigma_{T_2}^2},\\
\left. R_1+R_2\geq
I(Y_1,Y_2;W_1,W_2)=\frac{1}{2}\log\frac{(\sigma_X^2+\sigma_{N_1}^2+\sigma_{T_1}^2)
(\sigma_X^2+\sigma_{N_2}^2+\sigma_{T_2}^2)-\sigma_X^4}{\sigma_{T_1}^2\sigma_{T_2}^2}\right\}
\end{eqnarray*}
and
\begin{eqnarray*}
\Sigma''(D_3)=\left\{(\sigma_{T_1}^2,\sigma_{T_2}^2): D_3\geq
\mathbb{E}(X-\mathbb{E}(X|W_1,W_2))^2=
\left(\frac{1}{\sigma^2_X}+\frac{1}{\sigma_{N_1}^2+\sigma_{T_1}^2}+\frac{1}{\sigma_{N_2}^2+\sigma_{T_2}^2}\right)^{-1}\right\}.
\end{eqnarray*}

Let
$\mathcal{R}_{IPPR}(D_3)=\mathcal{R}_{IPPR}(\sigma^2_X,\sigma^2_X,D_3)$.
For comparison, we plot $\mathcal{R}_{IPPR}(D_3)$ and
$\mathcal{R}(D_3)$ in Fig.2. It's clear that
$\mathcal{R}_{IPPR}(D_3)\subsetneqq\mathcal{R}(D_3)$. A natural
question is to ask whether it is still possible to achieve not
only $D_3$ but also nontrivial $D_1$ and $D_2$ when the system is
operated in $\mathcal{R}(D_3)\cap\mathcal{R}_{IPPR}^c(D_3)$.

It has been shown in \cite{Viswanath}\cite{ChenCEO} that
$\mathcal{R}''(\sigma_{T_1}^2,\sigma_{T_2}^2)$ is a
contra-polymatroid. Its typical shape is plotted in Fig.3. The two
vertices $E_1$ and $E_2$ of
$\mathcal{R}''(\sigma_{T_1}^2,\sigma_{T_2}^2)$ are of special
importance, where
\begin{eqnarray*}
E_1&=&\left(I(Y_1;W_1),I(Y_2;W_2|W_1)\right)\\&=&\left(\frac{1}{2}\log\frac{\sigma_X^2+\sigma_{N_1}^2+\sigma_{T_1}^2}{\sigma_{T_1}^2},\frac{1}{2}\log\frac{(\sigma_X^2+\sigma_{N_1}^2+\sigma_{T_1}^2)
(\sigma_X^2+\sigma_{N_2}^2+\sigma_{T_2}^2)-\sigma_X^4}{\sigma_X^2\sigma_{T_1}^2+\sigma_{N_2}^2\sigma_{T_1}^2+\sigma_{T_1}^2\sigma_{T_2}^2}\right), \\
E_2&=&\left(I(Y_1;W_1|W_2),I(Y_2;W_2)\right)\\&=&\left(\frac{1}{2}\log\frac{(\sigma_X^2+\sigma_{N_1}^2+\sigma_{T_1}^2)
(\sigma_X^2+\sigma_{N_2}^2+\sigma_{T_2}^2)-\sigma_X^4}{\sigma_X^2\sigma_{T_2}^2+\sigma_{N_1}^2\sigma_{T_2}^2+\sigma_{T_1}^2\sigma_{T_2}^2},\frac{1}{2}\log\frac{\sigma_X^2+\sigma_{N_2}^2+\sigma_{T_2}^2}{\sigma_{T_2}^2}\right).
\end{eqnarray*}
Roughly speaking, the operational meaning for $E_1$ is that
encoder 1 employs the conventional lossy source coding and encoder
2 does the Wyner-Ziv coding; while for $E_2$, encoder 2 employs
the conventional lossy source coding and encoder 1 does the
Wyner-Ziv coding. The Wyner-Ziv coding requires random binning
scheme but the conventional lossy source coding does
not\footnote{For the conventional lossy source coding, the
codeword is directly revealed to the decoder, which corresponds to
the trivial binning scheme that each bin contains only one
codeword.}. So when the system is operated at $E_i$ $(i=1,2)$, the
decoder $i$ can decode the data sent by encoder $i$ and achieve
\begin{eqnarray}
D_i=\mathbb{E}(X-\mathbb{E}(X|W_1))^2=\left(\frac{1}{\sigma^2_X}+\frac{1}{\sigma^2_{N_i}+\sigma^2_{T_i}}\right)^{-1}.
\label{d_i}
\end{eqnarray}
Furthermore, for vertex $E_i$, we have
\begin{eqnarray}
R_i=I(Y_i;W_i)=\frac{1}{2}\log\left(\frac{\sigma_X^2+\sigma_{N_1}^2+\sigma_{T_1}^2}{{\sigma_{T_1}^2}}\right).
\label{r_i}
\end{eqnarray}
Combining (\ref{d_i}) and (\ref{r_i}), we get
$R_i=R(D_i,\sigma^2_{N_i})$. That is to say, the system can
achieve $D^*_i(R_i)$ at decoder $i$ when it is operated at $E_i$,
$i=1,2$. As shown in Fig. 3, $\mathcal{R}(D_3)$ is the union of
$\mathcal{R}''(\sigma_{T_1}^2,\sigma_{T_2}^2)$. The boundary of
$\mathcal{R}(D_3)$ can be divided into three pieces: $A, B$ and
$C$. Each point on $A$ corresponds to vertex $E_1$ of
$\mathcal{R}''(\sigma_{T_1}^2,\sigma_{T_2}^2)$ for some
$(\sigma_{T_1}^2,\sigma_{T_2}^2)$.  Each point on $C$ corresponds
to vertex $E_2$ of $\mathcal{R}''(\sigma_{T_1}^2,\sigma_{T_2}^2)$
for some $(\sigma_{T_1}^2,\sigma_{T_2}^2)$. So when the system is
operated at $(R_1,R_2)$ on curve $A$, it can achieve $D^*_1(R_1)$
at decoder 1 and at the same time achieve $D^*_3(R_1,R_2)$ at
decoder 3. Curve $C$ is similar to Curve $A$ with the only
difference that now the system can achieve $D^*_2(R_2)$ at decoder
2. This observation immediately yields the following partial
characterization of $\mathcal{Q}$:

Let $
D_{i,\min}=\mathbb{E}(X-\mathbb{E}(X|Y_i)))^2=(1/\sigma_X^2+1/\sigma_{N_i}^2)^{-1}$,
$i=1,2$, and
$D_{3,\min}=\mathbb{E}(X-\mathbb{E}(X|Y_1,Y_2)))^2=(1/\sigma_X^2+1/\sigma_{N_1}^2+1/\sigma_{N_1}^2)^{-1}$.

For any $D_3\in [D_{3,\min},\sigma^2_X]$, let
\begin{eqnarray*}
&\tilde{\sigma}_{T_1}^2=\left\{\begin{array}{ll}
  \left(\frac{\sigma_{N_1}^2}{D_{3,\min}}-\frac{\sigma_{N_1}^2}{D_3}\right)\left(\frac{1}{\sigma_{N_1}^2}-\frac{1}{D_{2,\min}}+\frac{1}{D_3}\right)^{-1}, \hspace{0.3in}\frac{2}{\max(\sigma^2_{N_1},\sigma^2_{N_2})}+\frac{1}{D_3}-\frac{1}{D_{3,\min}}\geq 0\\
  \left(\frac{\sigma_{N_1}^2}{D_{1,\min}}-\frac{\sigma_{N_1}^2}{D_3}\right)\left(\frac{1}{D_3}-\frac{1}{\sigma_X^2}\right)^{-1}, \hspace{0.5in}\frac{2}{\max(\sigma^2_{N_1},\sigma^2_{N_2})}+\frac{1}{D_3}-\frac{1}{D_{3,\min}}<0\mbox{ and }\sigma^2_{N_1}<\sigma^2_{N_2}\\
  \infty,  \hspace{2.8in}\mbox{otherwise}, \\
\end{array}\right.&\\
&\tilde{\sigma}_{T_2}^2=\left\{\begin{array}{ll}
  \left(\frac{\sigma_{N_2}^2}{D_{3,\min}}-\frac{\sigma_{N_2}^2}{D_3}\right)\left(\frac{1}{\sigma_{N_2}^2}-\frac{1}{D_{1,\min}}+\frac{1}{D_3}\right)^{-1}, \hspace{0.3in}\frac{2}{\max(\sigma^2_{N_1},\sigma^2_{N_2})}+\frac{1}{D_3}-\frac{1}{D_{3,min}}\geq 0\\
  \left(\frac{\sigma_{N_2}^2}{D_{2,\min}}-\frac{\sigma_{N_2}^2}{D_3}\right)\left(\frac{1}{D_3}-\frac{1}{\sigma_X^2}\right)^{-1}, \hspace{0.5in}\frac{2}{\max(\sigma^2_{N_1},\sigma^2_{N_2})}+\frac{1}{D_3}-\frac{1}{D_{3,\min}}<0\mbox{ and }\sigma^2_{N_1}>\sigma^2_{N_2}\\
  \infty,  \hspace{2.8in}\mbox{otherwise}.
\end{array}\right.&\\
\end{eqnarray*}
We have
\begin{itemize}
\item [(1)] For any $D_1\in
[D_{1,\min},\sigma_X^2(\sigma_{N_1}^2+\tilde{\sigma}_{T_1}^2)/(\sigma_X^2+\sigma_{N_1}^2+\tilde{\sigma}_{T_1}^2)]$,
\begin{eqnarray*}
\left\{(R_1,R_2):(R_1,R_2,D_1,\sigma^2_X,D_3)\in
\mathcal{Q}\right\}=\left\{(R_1,R_2):(R_1,R_2)\in\mathcal{R}(D_3),R_1\geq
R(D_1,\sigma_{N_1}^2)\right\}.
\end{eqnarray*}
\item [(2)] For any $D_2\in
[D_{2,\min},\sigma_X^2(\sigma_{N_2}^2+\tilde{\sigma}_{T_2}^2)/(\sigma_X^2+\sigma_{N_2}^2+\tilde{\sigma}_{T_2}^2)]$,
\begin{eqnarray*}
\left\{(R_1,R_2):(R_1,R_2,\sigma^2_X,D_2,D_3)\in
\mathcal{Q}\right\}=\left\{(R_1,R_2):(R_1,R_2)\in\mathcal{R}(D_3),R_2\geq
R(D_2,\sigma_{N_2}^2)\right\}.
\end{eqnarray*}
\item [(3)] For any $D_1\in
[D_{1,\min},\sigma_X^2(\sigma_{N_1}^2+\tilde{\sigma}_{T_1}^2)/(\sigma_X^2+\sigma_{N_1}^2+\tilde{\sigma}_{T_1}^2)]$
and $D_2\in
[D_{2,\min},\sigma_X^2(\sigma_{N_2}^2+\tilde{\sigma}_{T_2}^2)/(\sigma_X^2+\sigma_{N_2}^2+\tilde{\sigma}_{T_2}^2)]$,
\begin{eqnarray*}
\left\{(R_1,R_2):(R_1,R_2,D_1,D_2,D_3)\in
\mathcal{Q}\right\}=\left\{(R_1,R_2):R_i\geq
R(D_i,\sigma_{N_i}^2), i=1,2 \right\}.
\end{eqnarray*}
\end{itemize}

\begin{figure}[hbt]
\centering
\includegraphics[scale=0.5]{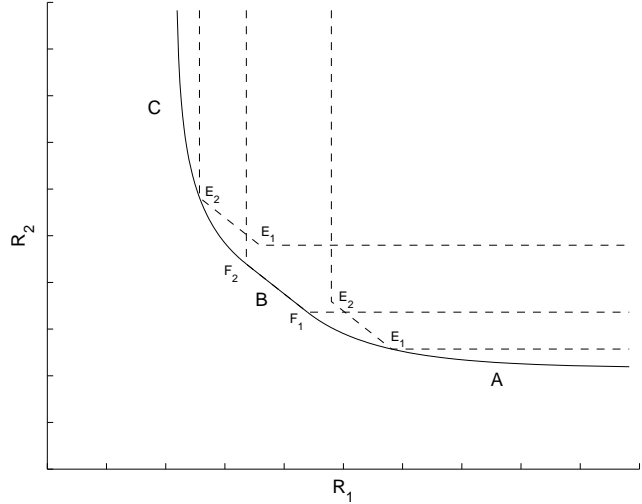}
\caption{The boundary of $\mathcal{R}(D_3)$}
\end{figure}

Since any rate tuple $(R_1,R_2)$ on line segment $B$ can be viewed
as the timesharing of $F_1$ and $F_2$, it implies that when the
system is operated on line segment $B$, it can achieve
$D^*_3(R_1,R_2)$ at decoder 3 and at the same time achieve
nontrivial $D_1$ and $D_2$ at decoder 1 and decoder 2,
respectively.

The IPPR scheme can achieve $D^*_1(R_1)$ and $D^*_2(R_2)$ at
decoder 1 and decoder 2 respectively, but can not achieve
$D^*_3(R_1,R_3)$ at decoder 3 in general  The coding scheme we
described above can achieve $D^*_3(R_1,R_2)$ at decoder 3 (at
least in the case of quadratic Gaussian CEO problem) and at the
same time achieve nontrivial $D_1$ and $D_2$ at decoder 1 and
decoder 2, but in general we have $D_i>D^*_i(R_i)$, $i=1,2$. We
will see that these are two extremes and there exist many other
schemes in between.

Like the CEO problem, the multiple description problem has been
studied for years and many multiple description coding schemes
have been proposed. Here we outline the common feature of the
existing multiple description coding schemes: encoder $i$
$(i=1,2)$, instead of sending an index $C^{(n)}_i$, sends a
vector, say $(C^{(n)}_{i,1},C^{(n)}_{i,2})$; decoder $i$ $(i=1,2)$
can only decode the $C^{(n)}_{i,1}-$part; decoder 3 can decode
both $(C^{(n)}_{1,1},C^{(n)}_{1,2})$ and
$(C^{(n)}_{2,1},C^{(n)}_{2,2})$. Clearly, this idea is also
applicable in the distributed source coding. Moreover, we can see
that the IPPR scheme corresponds to the case where $C^{(n)}_{1,2}$
and $C^{(n)}_{2,2}$ are constants.

In the next section, we propose a robust distributed coding scheme
by combining the random binning technique and the ideas from the
multiple description coding.

\section{Main Theorems}

\subsection{An Achievable Rate-Distortion Region}

\begin{theorem}\label{theorem1}
$(R_1,R_2,D_1,D_2,D_3)$ is achievable, if there exist random
variables $(U_1,U_2,W_1,W_2)$ jointly distributed with the generic
source variables $(X,Y_1,Y_2)$ such that the following properties
are satisfied:
\begin{itemize}
  \item [(i)] $(U_1,W_1)\rightarrow Y_1\rightarrow (X, Y_2, U_2,
  W_2)$ and $(U_2,W_2)\rightarrow Y_2\rightarrow (X, Y_1, U_1,
  W_1)$.
  \item [(ii)] $(R_1,R_2)\in\mathcal{R}(U_1,U_2,W_1,W_2)$, where
\begin{eqnarray*}
\mathcal{R}(U_1,U_2,W_1,W_2)=\left\{(R_1,R_2): R_1\geq
I(Y_1;U_1)+I(Y_1;W_1|U_1,U_2,W_2),\right.\\
R_2\geq I(Y_2;U_2)+I(Y_2;W_2|U_1,U_2,W_1),\\
\left. R_1+R_2\geq
I(Y_1;U_1)+I(Y_2;U_2)+I(Y_1,Y_2;W_1,W_2|U_1,U_2)\right\}.
\end{eqnarray*}
  \item [(iii)] There exist functions
$f_i:\mathcal{W}_i\rightarrow\mathcal{X}\ (i=1,2)$ and
$f_3:\mathcal{U}_1\times\mathcal{W}_1\times\mathcal{U}_2\times\mathcal{W}_2\rightarrow\mathcal{X}$
such that $\mathbb{E} d(X, {\hat X}_i)\leq D_i, i=1,2,3$, where
${\hat X}_1=f_1(U_1)$, ${\hat X}_2=f_2(U_2)$ and ${\hat
X}_3=f_3(U_1, W_1, U_2, W_2)$.
\end{itemize}

If $\mathcal{C}$ denotes the set of these achievable quintuples,
then time sharing yields that $\mathcal{Q}_{in}\triangleq
conv(\mathcal{C})$ is also an achievable region.
\end{theorem}

\begin{proof}
See Appendix I.
\end{proof}

Remark:
\begin{enumerate}
  \item Cardinality bound:
  By invoking the support
lemma \cite[pp.310]{CK}, $\mathcal{U}_1$ must have
$|\mathcal{Y}_1|-1$ letters to preserve the probability
distribution $P(y_1)$ and 5 more to preserve
$I(Y_1;U_1)+I(Y_1;W_1|U_1,U_2,W_2)$, $I(Y_2;W_2|U_1,U_2,W_1)$,
$I(Y_1;U_1)+I(Y_1,Y_2;W_1,W_2|U_1,U_2)$, $D_1$ and $D_3$, so
$|\mathcal{U}_1|=|\mathcal{Y}_1|+4$ suffices. $\mathcal{W}_1$ must
have $|\mathcal{Y}_1||\mathcal{U}_1|-1$ letters to preserve the
probability distribution $P(y_1,u_1)$ and 4 more to preserve
$I(Y_1;W_1|U_1,U_2,W_2)$, $I(Y_2;W_2|U_1,U_2,W_1)$,
$I(Y_1,Y_2;W_1,W_2|U_1,U_2)$ and $D_3$. Thus it suffices to have
$|\mathcal{W}_1|=|\mathcal{Y}_1||\mathcal{U}_1|-1+4=
|\mathcal{Y}_1|^2+4|\mathcal{Y}_1|+3$. Similarly, we have
$|\mathcal{U}_2|=|\mathcal{Y}_2|+4$, $|\mathcal{W}_2|=
|\mathcal{Y}_2|^2+4|\mathcal{Y}_2|+3$.
  \item It's easy to check that $\mathcal{R}(U_1,U_2,W_1,W_2)$
  is a contra-polymatroid. See \cite{Edmonds} for
  the defintion of contra-polymatroid and \cite{Viswanath,ChenCEO,TseCEO,Han,TsePolymatroid,SuccessiveCEO}
  for its applications in information
  theory.
  \item Let
  $W'_i=(U_i,W_i)$, $i=1,2$. It's
  easy to check that $\mathcal{R}(U_1,U_2,W'_1,W'_2)=\mathcal{R}(U_1,U_2,W_1,W_2)$ and $\hat
  X_3=f_3(U_1,U_2,W_1,W_2)=f'_3(W'_1,W'_2)$. So there is no loss of generality to assume $U_i\rightarrow W_i\rightarrow Y_i$ and define $f_3$
  on $\mathcal{W}_1\times\mathcal{W}_2$.
  \item Let $\mathcal{R}'(U_1,U_2,W_1,W_2)=\{(R_1,R_2):R_1\geq I(Y_1;U_1)+I(Y_1;W_1|U_1,U_2,W_2),R_2\geq
  I(Y_2;U_2)+I(Y_2;W_2|U_1,U_2)\}$, $\mathcal{R}''(U_1,U_2,W_1,W_2)=\{(R_1,R_2):R_1\geq I(Y_1;U_1)+I(Y_1;W_1|U_1,U_2),R_2\geq
  I(Y_2;U_2)+I(Y_2;W_2|U_1,U_2,W_1)\}$. Since
  $conv(\mathcal{R}'(U_1,U_2,W_1,W_2)\cup\mathcal{R}''(U_1,U_2,W_1,W_2))=\mathcal{R}(U_1,U_2,W_1,W_2)$,
  we can let
  $(R_1,R_2)\in\mathcal{R}'(U_1,U_2,W_1,W_2)\cup\mathcal{R}''(U_1,U_2,W_1,W_2)$
  in property (ii)
  without affecting $\mathcal{Q}_{in}$.
\end{enumerate}

A counter example constructed by K\"{o}rner and Marton
\cite{Marton} shows that $conv(\mathcal{C})\subsetneqq\mathcal{Q}$
in general. Actually even some special cases of our problem such
as the multiple description problem and the CEO problem are the
open problems of long standing. But for the following case, a
stronger assertion can be made.

\begin{Corollary}
For any $D_1$ and $D_3$, we have $\min\{R_1:\exists R_2 \mbox{
such that } (R_1,R_2,D_1,D_{\max},D_3)\in \mathcal{Q}
\}=\min[I(Y_1;U_1)+I(Y_1;W_1|Y_2,U_1)]$, where the minimization is
over the set of all random variables $(U_1,W_1)$ jointly
distributed with the generic source variables $(X,Y_1,Y_2)$ such
that the following conditions are satisfied:
\begin{itemize}
  \item [(i)] $(U_1,W_1)\rightarrow Y_1\rightarrow (X, Y_2)$.
  \item [(ii)] There exist functions  $f:
  \mathcal{U}_1\rightarrow\mathcal{X}$, $g:
  \mathcal{U}_1\times\mathcal{W}_1\times\mathcal{Y}_2\rightarrow\mathcal{X}$
such that $\mathbb{E}d(X, f(U_1))\leq D_1$, $\mathbb{E}d(X, g(Y_2,
U_1,W_1))\leq D_3$.
\item [(iii)] $|\mathcal{U}_1|=|\mathcal{Y}_1|+2$, $|\mathcal{W}_1|=(|\mathcal{Y}_1|+1)^2$.
\end{itemize}
Remark: For the same reason as before, there is no loss of
optimality to assume $U_1\rightarrow W_1\rightarrow Y_1$ and
define $g$ on $\mathcal{W}_1\times\mathcal{Y}_2$.

\begin{proof}
Since here we are only interested in minimizing $R_1$ under the
distortion constraints $D_1$ and $D_3$, there is no loss of
generality to assume that $R_2$ is large enough so that
$\{Y_2(t)\}_{t=1}^\infty$ can be recovered losslessly at decoder
3. In the case, our problem becomes the ``noisy" Heegard-Berger
problem. Its direct coding theorem can be easily reduced from
Theorem 1 while the converse coding theorem can be proved along
the same line as the converse in \cite{Heegard}.
\end{proof}
\end{Corollary}

\subsection{Distributed Source Coding with Identical Encoders}

For many applications, it is preferable to have encoders with
identical functionalities. It is thus interesting to study the
distributed source coding system with identical encoders. In order
to have $f^{(n)}_{E,1}=f^{(n)}_{E,2}$, two necessary conditions
are required: (1) $R_1=R_2$, (2) $\mathcal{Y}_1=\mathcal{Y}_2$. We
need the first condition to guarantee the range cardinalities of
$f^{(n)}_{E,1}$ and $f^{(n)}_{E,2}$ are the same, and the second
condition to guarantee these two encoding functions are defined on
the same domain. Without loss of generality, we can assume the
second condition is satisfied since we can let
$\mathcal{Y}=\mathcal{Y}_1\cup\mathcal{Y}_2$ and extend the
probability distribution $P(x, y_1, y_2)$ to be defined on
$\mathcal{X}\times \mathcal{Y}\times \mathcal{Y}$.

But even with these two conditions, it can not be guaranteed that
the resulting encoding functions $f^{(n)}_{E,1}$ and
$f^{(n)}_{E,2}$ in Theorem 1 are identical. Intuitively, in order
to minimize the distortion $D_3$ for the fixed rate constraints,
the $f^{(n)}_{E,1}$ and $f^{(n)}_{E,2}$ generated by the random
coding argument in the proof of Theorem 1 should be complementary
to each other instead of being identical. We may imagine that a
restriction on the identicalness of $f^{(n)}_{E,1}$ and
$f^{(n)}_{E,2}$ may incur performance loss. But we will show that
for many cases, no performance degradation will be caused.
Firstly, we need a formal definition.

\begin{definition}
The quadruple $(R, D_1, D_2, D_3)$ is called achievable with
identical encoders, if for any $\epsilon>0$, there exists an $n_0$
such that for all $n>n_0$ there exist an encoding function:
\begin{equation*}
  f_{E}^{(n)}:
  \mathcal{Y}^{n}\rightarrow\mathcal{C}^{(n)} \quad
  \log|\mathcal{C}^{(n)}|\leq n(R+\epsilon)
\end{equation*}
and decoders:
\begin{equation*}
f^{(n)}_{D, i}:\ \mathcal{C}^{(n)}\rightarrow \mathcal{X}^n \quad
i=1,2
\end{equation*}
\begin{equation*}
  f^{(n)}_{D, 3}:\ \mathcal{C}^{(n)}\times
\mathcal{C}^{(n)}\rightarrow \mathcal{X}^n
\end{equation*}
such that for $\hat X^n_i=f^{(n)}_{D,i}(f^{(n)}_{E}(Y^n_i)),
i=1,2$, and for $\hat
X^n_3=f^{(n)}_{D,3}(f^{(n)}_{E}(Y^n_1),f^{(n)}_{E}(Y^n_2))$,
\begin{equation*}
\frac{1}{n}E\sum\limits_{t=1}^n d(X(t), \hat X_i(t))<D_i+\epsilon
\quad i=1,2,3.
\end{equation*}
Let $\widetilde{\mathcal{Q}}$ denote the set of all achievable
quadruples.
\end{definition}

It is clear by definition that
$\min\{R:(R,D_1,D_2,D_3)\in\widetilde{\mathcal{Q}}\}$ is lower
bounded by $\min \{R:(R,R,D_1,D_2,D_3)\in\mathcal{Q}\}$. The
following theorem provides an upper bound on
$\min\{R:(R,D_1,D_2,D_3)\in\widetilde{\mathcal{Q}}\}$.

\begin{theorem}
For any feasible\footnote{We say $(D_1,D_2,D_3)$ is feasible if
$\{(H(Y_1),H(Y_2),D_1,D_2,D_3)\in\mathcal{Q}\}$.} $(D_1,D_2,D_3)$,
\begin{itemize}
\item [(1)] $\min\{R:(R,D_1,D_2,D_3)\in\widetilde{\mathcal{Q}}\}\leq
\min(\min\{R_1+R_2:(R_1,R_2,D_1,D_2,D_3)\in\mathcal{Q}\},H_{\max})$,
where $H_{\max}=\max(H(Y_1),H(Y_2))$.
\item [(2)]
$\min\{R:(R,D_1,D_2,D_3)\in\widetilde{\mathcal{Q}}\}=\min
\{R:(R,R,D_1,D_2,D_3)\in\mathcal{Q}\}$ if $P(Y_1=y)\neq P(Y_2=y)$
for some $y\in\mathcal{Y}$.
\end{itemize}
\begin{proof}
(1) For any $f^{(n)}_{E,1}$ and $f^{(n)}_{E,2}$ , let
$f^{(n)}_{E}=(f^{(n)}_{E,1},f^{(n)}_{E,2})$.  So if
$|f^{(n)}_{E,1}|=2^{nR_1},|f^{(n)}_{E,2}|=2^{nR_2}$, then we have
$|f^{(n)}_{E}|\leq 2^{n(R_1+R_2)}$. It's clear that if we replace
both $f^{(n)}_{E,1}$ and $f^{(n)}_{E,2}$ by $f^{(n)}_{E}$, no
additional estimation distortion will be incurred. Hence we have
$\min\{R:(R,D_1,D_2,D_3)\in\widetilde{\mathcal{Q}}\}\leq
\min\{R_1+R_2:(R_1,R_2,D_1,D_2,D_3)\in\mathcal{Q}\}$.

On the other hand, for any $\epsilon>0$, we can find a universal
lossless source encoding function with rate $R\leq
H_{\max}+\epsilon$ that works for both $\{Y_1(t)\}_{t=1}^\infty$
and $\{Y_2(t)\}_{t=1}^\infty$. This yields
$\min\{R:(R,D_1,D_2,D_3)\in\widetilde{\mathcal{Q}}\}\leq
H_{\max}$.

(2) The above proof essentially constructed a common encoder by
combining two encoding functions. We now show that if
$P(Y_1=y)\neq P(Y_2=y)$ for some $y\in\mathcal{Y}$, i.e.,
$\{Y_1(t)\}_{t=1}^\infty$ and $\{Y_2(t)\}_{t=1}^\infty$ are
distinguishable, we can combine two encoding functions in a more
efficient way.

For $\delta
>$ 0, let $T^n_{[Y_1]_\delta}$ be the set of $\delta$-typical
$Y_1$-vectors with length $n$. $T^n_{[Y_2]_\delta}$ is similarly
defined. If $P(Y_1=y)\neq P(Y_2=y)$ for some $y\in\mathcal{Y}$,
then $T^n_{[Y_1]_\delta}\bigcap T^n_{[Y_2]_\delta}=\emptyset$ when
$\delta$ is small enough. Note: here $\delta$ does not depend on
$n$. For any $\epsilon>0$ and any $R$ such that
$(R,R,D_1,D_2,D_3)\in\mathcal{Q}$, by Definition 1, there exist
two encoding functions: $f^{(n)}_{E,1}:\mathcal{Y}^n\rightarrow
\mathcal{C}^{(n)}_1$ and $f^{(n)}_{E,2}:\mathcal{Y}^n\rightarrow
\mathcal{C}^{(n)}_2$ with $(\log |\mathcal{C}^{(n)}_i|)/n\leq
R+\epsilon$, $i=1,2$. Here we make $n$ arbitrarily large via
concatenation. Without loss of generality, we assume
$\mathcal{C}^{(n)}_{1}=\mathcal{C}^{(n)}_{2}=\mathcal{C}^{(n)}$
and $(\log|\mathcal{C}^{(n)}|)/n=R+\epsilon$. Define $f^{(n)}_E:
\mathcal{Y}^n\rightarrow \mathcal{C}^{(n)}$ such that
\begin{eqnarray*}
f^{(n)}_{E}(Y^n)=&\left\{\begin{array}{c}
  f^{(n)}_{E,1}(Y^n),  \hspace{0.3in}   Y^n\in T^n_{[Y_1]_\delta}\\
  f^{(n)}_{E,2}(Y^n),   \hspace{0.3in}  Y^n\notin T^n_{[Y_1]_\delta}
\end{array}\right.&.
\end{eqnarray*}

Since $P(Y^{n}_1\not\in T^n_{[Y_1]_\delta} \mbox{ or }
Y^{n}_2\not\in T^n_{[Y_2]_\delta})\leq P(Y^{n}_1\not\in
T^n_{[Y_1]_\delta})+P(Y^{n}_2\not\in
T^n_{[Y_2]_\delta})=\epsilon(n)\rightarrow 0$ as
$n\rightarrow\infty$, if we replace both $f^{(n)}_{E,1}$ and
$f^{(n)}_{E,2}$ by $f^{(n)}_{E}$, the additional estimation
distortion it may incur is at most $\epsilon(n)d_{\max}$, which is
negligible when $n$ is large enough.
\end{proof}
\end{theorem}

The above proof essentially suggests a way to convert a
distributed source coding system with different encoders to a
system with identical encoders. We can conclude that for a
distributed source coding system, we can use identical
encoders\footnote{possibly at the price of high complexity.} and
still achieve optimal rate-distortion tradeoff when the marginal
distributions of the observations are different. But If
$P(Y_1=y)=P(Y_2=y)$ for all $y\in\mathcal{Y}$, then the
restriction $f^{(n)}_{E,1}=f^{(n)}_{E,2}$ will cause performance
loss in general. The simplest example is to set $Y_1=Y_2=X$. Now
if we let $f^{(n)}_{E,1}=f^{(n)}_{E,2}$, then no diversity gain
can be achieved at decoder 3.

\subsection{Multiple Description with Noisy Observations}

If there exist $f_1$ and $f_2$ such that $Y=f_1(Y_1)=f_2(Y_2)$
with probability one and $X\rightarrow Y\rightarrow (Y_1,Y_2)$ ,
our problem becomes the multiple description problem with noisy
observations. In this case, we can directly adopt the multiple
description coding scheme with only a slight change.

\begin{theorem}\label{theorem3}
\begin{itemize}
 \item [(1)]$(R_1,R_2,D_1,D_2,D_3)$ is achievable if there exist random
variables $({\hat X}_0,{\hat X}_1,{\hat X}_2,{\hat X}_3)$ jointly
distributed with the generic source variables $(X,Y)$ such that
the following properties are satisfied:
\begin{itemize}
  \item [(i)] $X\rightarrow Y\rightarrow ({\hat
X}_0, {\hat X}_1, {\hat X}_2, {\hat X}_3)$,
\item [(ii)]
$R_1+R_2\geq 2I(Y;{\hat X}_0)+I({\hat X}_1;{\hat X}_2|{\hat
X}_0)+I(Y;{\hat X}_1,{\hat X}_2,{\hat X}_3|{\hat X}_0)$, $R_i\geq
I(Y;{\hat X}_0,{\hat X}_i)$, $i=1,2$.
  \item [(iii)] $E d(X, {\hat X}_i)\leq D_i, i=1,2,3$.
\end{itemize}
If $\mathcal{C'}$ denotes the set of these achievable quintuples,
then time sharing yields that $conv(\mathcal{C'})$ is also an
achievable region.
\item [(2)] Let $\mathcal{C}^*$ denote the subset of $\mathcal{C'}$ containing all those quintuples satisfying
(i)-(iii), with the additional conditions that (a) ${\hat X}_1$
and ${\hat X}_2$ are independent, (b) ${\hat X}_0$ is a constant.
Let
\begin{equation*}
R^*(D)=\min\limits_{\substack{\hat X:X\rightarrow Y\rightarrow
\hat X, \\Ed(X,\hat X)\leq D}} I(Y;\hat X),
\end{equation*}
which is the rate-distortion function with noisy observations
\cite{ZivWolf}\cite{BergerText}. Let
\begin{eqnarray*}
\mathcal{Q}(D_3)&=&\{(R_1,R_2,D_1,D_2,D_3)\in
\mathcal{Q}:R_1+R_2=R^*(D_3)\},
\\conv(\mathcal{C'})(D_3)&=&\{(R_1,R_2,D_1,D_2,D_3)\in
conv(\mathcal{C'}):R_1+R_2=R^*(D_3)\}.
\end{eqnarray*}
We have
\begin{equation*}
\mathcal{Q}(D_3)=conv(\mathcal{C'})(D_3)
\end{equation*}
\end{itemize}
\end{theorem}

\begin{proof}
Part (1) of the theorem follows from Markov lemma and Theorem 1
(specialized to 2-encoder case) in \cite{ref33}. Part (2) of the
theorem can be proved via a "continuity" argument similar to that
of \cite{Ahlswede} by replacing Shannon's rate distortion function
$R(D_3)$ with $R^*(D_3)$ and noticing the following Markov
relation: $X(t)\rightarrow Y(t)\rightarrow (Y_1^n,
Y_2^n)\rightarrow (f^{(n)}_{E,1}(Y_1^n),f^{(n)}_{E,2}(Y_2^n))
\rightarrow (\hat X_1(t),\hat X_2(t),\hat X_3(t))$.
\end{proof}

Theorem \ref{theorem1} is associated with a distributed source
coding scheme while Theorem \ref{theorem3} is associated with a
centralized source coding scheme. Here "distributed" and
"centralized" are in the statistical sense instead of geographical
sense. Even for the centralized coding scheme, we can put two
encoders as far as possible as long as long as the inputs of these
two encoders are the same. Since these two encoders have the same
inputs, one knows exactly the operation the other will take and
thus they can have arbitrary cooperation. In this sense, the
encoders in a centralized coding system should be viewed as the
different functionalities of a single encoder, no matter how far
away they are separated. For a distributed coding system, since
two encoders have different inputs, one does not know for sure
about the operation the other will take. Hence, the types of
cooperation between two encoders in a statistically distributed
system are very limited. On the other hand, since centralized
coding system is a special case of distributed coding system, one
would expect a unified approach to both of them. But it is easy
check that Theorem \ref{theorem1}, when particularized to the
centralized case (i.e., $Y_1=Y_2=Y$ with probability one), does
not coincide with Theorem \ref{theorem3}. That is to say, Theorem
\ref{theorem3} is not a ``centralized'' version of Theorem
\ref{theorem1}. Now a natural question arises: Does there exist a
distributed source coding scheme which subsumes the centralized
source coding scheme in Theorem \ref{theorem3} as a special case ?

Now we suggest a unified approach which incorporates these two
schemes in a single framework. The main ingredient is a concept
called the common part(/information) of two dependent random
variables in the sense of Gacs and K\"{o}rner \cite{Gacs} and
Witsenhausen \cite{WitCommon}. The following definition is from
\cite{CoverMul}.

\begin{definition}
The common part $Z$ of two random variables $Y_1$ and $Y_2$ is
defined by finding the maximum integer $k$ such that there exist
functions $f: \mathcal{Y}_1\rightarrow \{1,2,\cdots,k\}$ and $g:
\mathcal{Y}_2\rightarrow \{1,2,\cdots,k\}$
with $P(f(Y_1)=i)>0, P(g(Y_2)=i)>0, i=1,2,\cdots,k$, such that
$f(Y_1)=g(Y_2)$ with probability one and then defining
$Z=f(Y_1)\mbox{ }(=g(Y_2))$.
\end{definition}

With this definition, it is obvious that encoder 1 and encoder 2
can agree on the value of $Z$ with probability one. Therefore,
they can use efficient centralized coding scheme (of Theorem 3
type) for the common part $Z$ and then superimpose a distributed
coding scheme (of Theorem 1 type). This observation immediately
leads to the following theorem.

\begin{theorem}\label{theorem4}
Let $Z$ be the common part of $Y_1$ and $Y_2$.
$(R_1,R_2,D_1,D_2,D_3)$ is achievable if there exist random
variables $(U_1,U_2,W_1,W_2,Z_0,Z_1,Z_2,Z_3)$ jointly distributed
with the generic source variables $(X,Y_1,Y_2,Z)$ such that the
following properties are satisfied:
\begin{itemize}
  \item [(i)] $(X,Y_1,Y_2)\rightarrow Z\rightarrow
  (Z_0,Z_1,Z_2)$;
  \item [(ii)] $U_1\rightarrow (Y_1,Z_0,Z_1)\rightarrow (X, Y_2, Z_2, U_2)$ and $U_2\rightarrow (Y_1,Z_0,Z_2)\rightarrow (X, Y_1, Z_1,
  U_1)$;
  \item [(iii)] $Z_3\rightarrow (Z,Z_0,Z_1,Z_2)\rightarrow
  (X,Y_1,Y_2,U_1,U_2)$;
  \item [(iv)] $W_1\rightarrow (Y_1,Z_0,Z_1,Z_2,Z_3,U_1)\rightarrow
  (X,Y_2,U_2,W_2)$ and $W_2\rightarrow (Y_2,Z_0,Z_1,Z_2,Z_3,U_2)\rightarrow
  (X,Y_1,U_1,W_1)$;
  \item [(v)]
\begin{eqnarray*}
R_1&\geq&
I(Y_1;Z_0,Z_1,U_1)+I(Y_1;W_1|Z_0,Z_1,Z_2,Z_3,U_1,U_2,W_2)
\\R_2&\geq&
I(Y_2;Z_0,Z_2,U_2)+I(Y_2;W_2|Z_0,Z_1,Z_2,Z_3,U_1,U_2,W_1)\\R_1+R_2&\geq&
I(Y_1;Z_0,Z_1,U_1)+I(Y_2;Z_0,Z_2,U_2)+I(Z_1;Z_2|Z_0)\\&&+I(Z;Z_1,Z_2,Z_3|Z_0)+I(Y_1,Y_2;W_1,W_2|Z_0,Z_1,Z_2,Z_3,U_1,U_2).
\end{eqnarray*}
  \item [(iv)] There exist functions:
$f_i: \mathcal{U}_i\rightarrow\mathcal{X}$, $i=1,2$, and $f_3:
\mathcal{U}_1\times\mathcal{W}_1\times\mathcal{U}_2\times\mathcal{W}_2\rightarrow\mathcal{X}$
such that $E d(X, {\hat X}_i)\leq D_i$, $i=1,2,3$,  where ${\hat
X}_1=f_1(U_1)$, ${\hat X}_2=f_2(U_2)$ and ${\hat X}_3=f_3(U_1,
W_1, U_2, W_2)$.
\end{itemize}

If $\mathcal{C}''$ denotes the set of these achievable quintuples,
then time sharing yields that $conv(\mathcal{C}'')$ is also an
achievable region.
\end{theorem}
\begin{proof}
The proof is omitted since it's a straightforward combination of
Theorem \ref{theorem1} and Theorem \ref{theorem3}.
\end{proof}
Remark:
\begin{enumerate}
  \item Theorem \ref{theorem4} can be reduced to Theorem \ref{theorem1} by letting
  $(Z_0,Z_1,Z_2,Z_3)=\mbox{constant}$.
  If $X\rightarrow Z\rightarrow (Y_1,Y_2)$, then Theorem \ref{theorem4} can be specialized
  to Theorem \ref{theorem3} by setting
  $(U_1,U_2,W_1,W_2)=\mbox{constant}$
  and noticing there is no loss of generality to let $Z_1,Z_2,Z_3$ assume
  values in $\mathcal{X}$.
  \item The conventional distributed source coding scheme
  \cite{Berger}\cite{Tung} does not consider the common part (even
  it does exist) of the observations and thus requires very restricted long Markov chain conditions on the auxiliary random
  variables. As we have seen in Theorem \ref{theorem4}, the long Markov chain conditions are not
  always necessary, at least in the case when there exists a common
  part in two observations.
  \item Theorem \ref{theorem4} essentially suggests an approach to bridging the distributed
  source coding scheme and the centralized source coding scheme.
  But for many cases, no common part exists for $Y_1$ and $Y_2$ even when they are highly
  correlated. Hence it is of special interest to see whether there
  exists a general coding scheme that can transit smoothly from a distributed
  scheme to a centralized scheme when $Y_1$ and
  $Y_2$ become more and more correlated but no common part exists.
\end{enumerate}

\section{Gaussian Case}

In this section, we apply the general results obtained in the
previous section to analyze the Gaussian case with squared
distortion measure. Although most of the results in Section IV are
proved for the finite alphabet case with bounded distortion
measure, they can be extended to the Gaussian case with squared
distortion measure by standard techniques
\cite{OohamaGaussian}\cite{WynerInfocontr}.

Let $\{X(t), Y_1(t)=X(t)+N_1(t),
Y_2(t)=X(t)+N_2(t)\}_{t=1}^{\infty}$  be i.i.d. zero-mean Gaussian
vectors such that $X(t)$, $N_1(t)$ and $N_2(t)$ are independent
with variances $\sigma^2_X, \sigma^2_{N_1}$ and $\sigma^2_{N_2}$
respectively. Without loss of generality, we only study the region
$\{(R_1,R_2,D_1,D_2,D_3)\in\mathcal{Q}: D_3\leq\min(D_1,D_2),
D_{i,\min}\leq D_i\leq\sigma_X^2, i=1,2,3\}$. For convenience, we
shall abuse the notation and denote this region by $\mathcal{Q}$.

\subsection{An Inner Bound of the Rate Distortion Region}

We derive the inner bound of the rate distortion region for the
Gaussian case by evaluating Thereom \ref{theorem1}. Let $W_1, U_1,
W_2, U_2$ be the auxiliary random variables jointly distributed
with the generic source variables $X, Y_1, Y_2$ such that
\begin{eqnarray*}
&\left\{\begin{array}{c}
  U_1=Y_1+T_{11} \\
  U_2=Y_2+T_{21}
\end{array}\right.\quad
\left\{\begin{array}{c}
  W_1=Y_1+T_{12} \\
  W_2=Y_2+T_{22}
\end{array}\right.&.
\label{equationWUY}
\end{eqnarray*}
Here $T_{11}, T_{12}, T_{21}, T_{22}$ are zero-mean Gaussian
random variables with variances $\sigma^2_{T_{11}},
\sigma^2_{T_{12}}, \sigma^2_{T_{21}}, \sigma^2_{T_{22}}$
respectively, and they are independent of $X, Y_1, Y_2$. Moreover,
$T_{11}, T_{12}$ are independent of $T_{21}, T_{22}$. The
correlation coefficient of $T_{i1}$ and $T_{i2}$ is $\rho_{T_i},
i=1,2$.

Let $W^*_i=\mathbb{E}(Y_i|U_i,W_i)$, $i=1,2$. It is easy to verify
that
\begin{eqnarray*}
&&\mathcal{R}(U_1,U_2,W_1,W_2)=\mathcal{R}(U_1,U_2,W^*_1,W^*_2),\\
&&\mathbb{E}(X-\mathbb{E}(X|W_1,W_2,U_1,U_2))^2=\mathbb{E}(X-\mathbb{E}(X|W^*_1,W^*_2,U_1,U_2))^2=\mathbb{E}(X-\mathbb{E}(X|W_1^*,W_2^*))^2.
\end{eqnarray*}
So there is no loss of generality to assume $U_i\rightarrow
W_i\rightarrow Y_i$, $i=1,2$, i.e., we can assume
$T_{i1}=T_{i2}+\Delta T_i$, $i=1,2$, where $\Delta
T_1\sim\mathcal{N}(0,\sigma^2_{T_{11}}-\sigma^2_{T_{12}})$ and
$\Delta T_1\sim\mathcal{N}(0,\sigma^2_{T_{21}}-\sigma^2_{T_{22}})$
are mutually independent, and they are independent of $X, Y_1,
Y_2, T_{12}$ and $T_{22}$.

Now by evaluating Theorem \ref{theorem1}, we get the following
achievable rate distortion region:
\begin{equation*}
  \mathcal{Q}_{in}=conv\left(\bigcup\limits_{(\sigma_{T_{11}}^2\geq\sigma_{T_{12}}^2,\sigma_{T_{21}}^2\geq\sigma_{T_{22}}^2)}\mathcal{C}(\sigma_{T_{11}}^2,\sigma_{T_{12}}^2,\sigma_{T_{21}}^2,\sigma_{T_{22}}^2)\right)
\end{equation*}
where
\begin{eqnarray*}
\mathcal{C}(\sigma_{T_{11}}^2,\sigma_{T_{12}}^2,\sigma_{T_{21}}^2,\sigma_{T_{22}}^2)\triangleq\left\{(R_1,R_2,D_1,D_2,D_3):
\frac{1}{D_i}\leq\frac{1}{\sigma_X^2}+\frac{1}{\sigma^2_{N_i}+\sigma^2_{T_{i1}}}, i=1,2,\right.\\
\frac{1}{D_3}\geq\frac{1}{\sigma_X^2}+\frac{1}{\sigma^2_{N_1}+\sigma^2_{T_{12}}}+\frac{1}{\sigma^2_{N_2}+\sigma^2_{T_{22}}},\\
R_1\geq\frac{1}{2}\log\frac{\sigma^2_{U_1}(\sigma^2_{W_{1}}\sigma^2_{W_{2}}-\sigma^4_X)}{\sigma^2_{T_{12}}(\sigma^2_{U_{1}}\sigma^2_{W_{2}}-\sigma^4_X)},\\
R_2\geq\frac{1}{2}\log\frac{\sigma^2_{U_{2}}(\sigma^2_{W_{1}}\sigma^2_{W_{2}}-\sigma^4_X)}{\sigma^2_{T_{22}}(\sigma^2_{W_{1}}\sigma^2_{U_{2}}-\sigma^4_X)},\\
\left.
R_1+R_2\geq\frac{1}{2}\log\frac{\sigma^2_{U_{1}}\sigma^2_{U_{2}}(\sigma^2_{W_{1}}\sigma^2_{W_{2}}-\sigma^4_X)}{\sigma^2_{T_{12}}\sigma^2_{T_{22}}(\sigma^2_{U_{1}}\sigma^2_{U_{2}}-\sigma^4_X)}\right\}
\end{eqnarray*}
and
$\sigma^2_{U_i}=\sigma^2_{X}+\sigma^2_{N_i}+\sigma^2_{T_{i1}}$,
$\sigma^2_{W_i}=\sigma^2_{X}+\sigma^2_{N_i}+\sigma^2_{T_{i2}}$,
$i=1,2$.

By Remark 4) of Theorem \ref{theorem1}, We can write
\begin{eqnarray*}
\mathcal{Q}_{in}=conv\left(\bigcup\limits_{(\sigma_{T_{11}}^2\geq\sigma_{T_{12}}^2,\sigma_{T_{21}}^2\geq\sigma_{T_{22}}^2)}\left(\mathcal{C}_1
(\sigma_{T_{11}}^2,\sigma_{T_{12}}^2,\sigma_{T_{21}}^2,\sigma_{T_{22}}^2)\cup\mathcal{C}_2
(\sigma_{T_{11}}^2,\sigma_{T_{12}}^2,\sigma_{T_{21}}^2,\sigma_{T_{22}}^2)\right)\right),
\end{eqnarray*}
where
\begin{eqnarray*}
\mathcal{C}_1(\sigma_{T_{11}}^2,\sigma_{T_{12}}^2,\sigma_{T_{21}}^2,\sigma_{T_{22}}^2)\triangleq\left\{(R_1,R_2,D_1,D_2,D_3):
\frac{1}{D_i}\leq\frac{1}{\sigma_X^2}+\frac{1}{\sigma^2_{N_i}+\sigma^2_{T_{i1}}}, i=1,2,\right.\\
\frac{1}{D_3}\geq\frac{1}{\sigma_X^2}+\frac{1}{\sigma^2_{N_1}+\sigma^2_{T_{12}}}+\frac{1}{\sigma^2_{N_2}+\sigma^2_{T_{22}}},\\
\left.
R_1\geq\frac{1}{2}\log\frac{\sigma^2_{U_1}(\sigma^2_{W_{1}}\sigma^2_{W_{2}}-\sigma^4_X)}{\sigma^2_{T_{12}}(\sigma^2_{U_{1}}\sigma^2_{W_{2}}-\sigma^4_X)},
 R_2\geq\frac{1}{2}\log\frac{\sigma^2_{U_{2}}(\sigma^2_{U_{1}}\sigma^2_{W_{2}}-\sigma^4_X)}{\sigma^2_{T_{22}}(\sigma^2_{U_{1}}\sigma^2_{U_{2}}-\sigma^4_X)}\right\}
\end{eqnarray*}
and
\begin{eqnarray*}
\mathcal{C}(\sigma_{T_{11}}^2,\sigma_{T_{12}}^2,\sigma_{T_{21}}^2,\sigma_{T_{22}}^2)\triangleq\left\{(R_1,R_2,D_1,D_2,D_3):
\frac{1}{D_i}\leq\frac{1}{\sigma_X^2}+\frac{1}{\sigma^2_{N_i}+\sigma^2_{T_{i1}}}, i=1,2,\right.\\
\frac{1}{D_3}\geq\frac{1}{\sigma_X^2}+\frac{1}{\sigma^2_{N_1}+\sigma^2_{T_{12}}}+\frac{1}{\sigma^2_{N_2}+\sigma^2_{T_{22}}},\\
\left.
R_1\geq\frac{1}{2}\log\frac{\sigma^2_{U_1}(\sigma^2_{W_{1}}\sigma^2_{U_{2}}-\sigma^4_X)}{\sigma^2_{T_{12}}(\sigma^2_{U_{1}}\sigma^2_{U_{2}}-\sigma^4_X)},
R_2\geq\frac{1}{2}\log\frac{\sigma^2_{U_{2}}(\sigma^2_{W_{1}}\sigma^2_{W_{2}}-\sigma^4_X)}{\sigma^2_{T_{22}}(\sigma^2_{W_{1}}\sigma^2_{U_{2}}-\sigma^4_X)}\right\}.
\end{eqnarray*}

\subsection{An Outer Bound of the Rate Distortion Region}
Let $\theta(t)=X(t)-S(t)$, $t=1,2,\cdots$, where
\begin{eqnarray*}
S(t)=\mathbb{E}(X(t)|Y_1(t),Y_2(t))=\frac{D_{3,\min}}{\sigma^2_{N_1}}Y_1(t)+\frac{D_{3,\min}}{\sigma^2_{N_2}}Y_2(t).
\end{eqnarray*}
$\theta(t)$ is Gaussian with mean 0 and variance $D_{3,\min}$, and
is independent of $Y_1(t)$ and $Y_2(t)$. Let
$d_X=\sigma^2_X-D_{3,\min}$ and $d_i=D_i-D_{3,\min}$, $i=1,2,3$.
Define
\begin{equation*}
  \mathcal{Q}_{out}=\bigcup\limits_{(r_{11},r_{12},r_{21},r_{22})\in\Sigma_{out}}\mathcal{C}_{out}(r_{11},r_{12},r_{21},r_{22})
\end{equation*}
where
\begin{eqnarray*}
\mathcal{C}_{out}(r_{11},r_{12},r_{21},r_{22})\triangleq\left\{(R_1,R_2,D_1,D_2,D_3): \frac{1}{D_i}\leq\frac{\exp(2r_{i1})}{\sigma_X^2}, R_i\geq r_{i1}+r_{i2}, i=1,2,\right.\\
\frac{1}{D_3}\leq \frac{1}{\sigma_X^2}+\frac{1-\exp(-2r_{12})}{\sigma_{N_1}^2}+\frac{1-\exp(-2r_{22})}{\sigma_{N_2}^2}\\
\left.
r_{11}+r_{21}\geq\frac{1}{2}\log\frac{\sigma_X^2}{D_3}+\lambda(D_1,D_2,D_3,r_{21},r_{22})\right\},
\end{eqnarray*}
\begin{eqnarray*}
&\lambda(D_1,D_2,D_3,r_{21},r_{22})=\left\{\begin{array}{ll}
  0, \hspace{2.55in} \zeta\leq d_1+d_2-d_X\\
  \frac{1}{2}\log\frac{d_X \zeta}{d_1d_2}, \hspace{2.01in}\zeta\geq \left(\frac{1}{d_1}+\frac{1}{d_2}-\frac{1}{d_X}\right)^{-1}\\
  \frac{1}{2}\log\frac{(d_X-\zeta)^2}
{(d_X-\zeta)^2-[\sqrt{(d_X-d_1)(d_X-d_2)}-\sqrt{(d_1-\zeta)(d_2-\zeta)}]^2},  \mbox{ otherwise},  \\
\end{array}\right.&\\
\end{eqnarray*}
\begin{equation*}
\hspace{-3.2in}\zeta=D_3D_{3,\min}\left(\frac{\exp(-2r_{21})}{\sigma^2_{N_1}}+\frac{\exp(-2r_{22})}{\sigma^2_{N_2}}\right)
\end{equation*}
and
\begin{eqnarray*}
\Sigma_{out}=\left\{(r_{11},r_{12},r_{21},r_{22})\in\mathbb{R}^4_+:
\frac{1}{\sigma_X^2}\exp(2r_{i1})\leq\frac{1}{\sigma_X^2}+\frac{1-\exp(-2r_{i2})}{\sigma_{N_i}^2},
i=1,2\right\}
\end{eqnarray*}

\begin{theorem}
$\mathcal{Q}\subseteq\mathcal{Q}_{out}$.
\end{theorem}
\begin{proof}
The proof is left to Appendix II.
\end{proof}

It's easy to check that $\{(R_1,R_2,D_3):
(R_1,R_2,\sigma^2_X,\sigma^2_X,D_3)\in\mathcal{Q}_{out}\}$ is the
rate distortion region of the quadratic Gaussian CEO problem
\cite{TseCEO} and $\mathcal{Q}_{out}$ converges to the rate
distortion region of the Gaussian multiple description problem
\cite{Ozarow} as $\sigma_{N_1}^2,\sigma_{N_2}^2\rightarrow 0.$
Moreover, $\mathcal{Q}_{in}$ and $\mathcal{Q}_{out}$ coincide in
some subregions as shown in the following corollary.

\begin{Corollary}
For any $(R_1,R_2,D_1,D_2,D_3)\in\mathcal{Q}$, if
$R_i=R(D_i,\sigma^2_{N_i})$, $i=1,2$, then $1/{D_3}\leq
1/{D_1}+1/{D_2}-1/{\sigma_X^2}$.
\end{Corollary}
\begin{proof}
From the outer bound, we have $R_i\geq r_{i1}+r_{i2}$,
$D_i\geq\sigma_X^2\exp(-2r_{i1})$, and
$1/{\sigma_X^2}+(1-\exp(-2r_{i2}))/{\sigma_{N_i}^2}\geq
\exp(2r_{i1})/\sigma_X^2$, $i=1,2$. Thus if
$R_i=R(D_i,\sigma^2_{N_i})$, $i=1,2$, then
\begin{eqnarray*}
r_{i1}&=&\frac{1}{2}\log\frac{\sigma_X^2}{D_i}, \\
r_{i2}&=&\frac{1}{2}\log\frac{\sigma_X^2D_i}{D_i\sigma_X^2-\sigma_X^2\sigma_{N_i}^2+D_i\sigma_{N_i}^2}.
\end{eqnarray*}
Therefore, we have
\begin{eqnarray*}
\frac{1}{D_3}&\leq&\frac{1}{\sigma_X^2}+\frac{1-\exp(-2r_{12})}{\sigma_{N_1}^2}+\frac{1-\exp(-2r_{22})}{\sigma_{N_2}^2}\\
&=&\frac{1}{D_1}+\frac{1}{D_2}-\frac{1}{\sigma_X^2}.
\end{eqnarray*}
\end{proof}

It's easy to check that IPPR scheme can achieve all the
$\left(R_1,R_2,D_1,D_2,D_3\right)$ satisfying
$R_i=R(D_i,\sigma^2_{N_i})$, $i=1,2$, and $1/{D_3}\leq
1/{D_1}+1/{D_2}-1/{\sigma_X^2}$. Hence for the quadratic Gaussian
case, it's impossible to have $D_3$ smaller than that achieved by
the IPPR scheme if decoder 1 and decoder 2 are rate-distortion
optimal.

\subsection{A Symmetric Case}
Let $\sigma^2_{N_1}=\sigma^2_{N_2}=\sigma^2_{N}$. It was computed
in \cite{ChenCEO} that
\begin{equation*}
R=\frac{1}{4}\log\left(\frac{\sigma^2_X}{D^*_3(R,R)}\left(\frac{2D^*_3(R,R)\sigma^2_X}{2D^*_3(R,R)\sigma^2_X-\sigma^2_X\sigma^2_N+D^*_3(R,R)\sigma^2_N}\right)^2\right),
\end{equation*}
or equivalently,
\begin{equation*}
D^*_3(R,R)=\frac{2\sigma^4_X\sigma^2_N+\sigma^2_X\sigma^4_N+2\sigma^6_X\exp(-4R)+2\sigma^4_X\exp(-2R)\sqrt{\sigma^4_X\exp(-4R)+2\sigma^2_X\sigma^2_N+\sigma^4_N}}{(2\sigma^2_X+\sigma^2_N)^2}.
\end{equation*}

Define
$\mathcal{D}_{12}(R)=\{(D_1,D_2):(R,R,D_1,D_2,D^*_3(R,R))\in\mathcal{Q}\}$.
We are not able to give a complete characterization of
$\mathcal{D}_{12}(R)$. Instead, we shall establish an inner bound
and an outer bound. Let
\begin{eqnarray*}
\mathcal{D}^{in}_{12}(R)=\{(D_1,D_2):(R,R,D_1,D_2,D^*_3(R,R))\in\mathcal{Q}_{in}\}
\end{eqnarray*}
and
\begin{eqnarray*}
\mathcal{D}^{out}_{12}(R)=\{(D_1,D_2):(R,R,D_1,D_2,D^*_3(R,R))\in\mathcal{Q}_{out}\}.
\end{eqnarray*}
Since
$\mathcal{Q}_{in}\subseteq\mathcal{Q}\subseteq\mathcal{Q}_{out}$,
it follows that
$\mathcal{D}^{in}_{12}(R)\subseteq\mathcal{D}_{12}(R)\subseteq\mathcal{D}^{out}_{12}(R)$.
Now we proceed to compute the explicit expressions of
$\mathcal{D}^{in}_{12}(R)$ and $\mathcal{D}^{out}_{12}(R)$.

It can be seen that $(R,R,D_1,D_2,D^*_3(R,R))\in\mathcal{Q}_{out}$
if and only if there exist
$(r_{11},r_{12},r_{21},r_{22})\in\mathbb{R}^4_+$ such that the
following set of inequalities are satisfied:
\begin{eqnarray}
R&\geq&r_{i1}+r_{i2},\label{c01}\\
D_i&\geq&\sigma^2_X\exp(-2r_{i1}),\label{c02}\\
\exp(2r_{i1})&\leq&1+\frac{\sigma^2_X-\sigma^2_X\exp(-2r_{i2})}{\sigma^2_{N_i}},\quad i=1,2,\label{c03}\\
r_{11}+r_{21}&\geq&\frac{1}{2}\log\frac{\sigma^2_X}{D^*_3(R,R)}+\lambda(D_1,D_2,D_3,r_{21},r_{22}),\label{c04}\\
\frac{1}{D^*_3(R,R)}&\leq&\frac{1}{\sigma^2_X}+\frac{1-2\exp(-2r_{12})}{\sigma^2_N}+\frac{1-2\exp(-r_{22})}{\sigma^2_N}.
\end{eqnarray}
By \cite[Lemma 3.5]{SuccessiveCEO}, if
$(r_{11},r_{12},r_{21},r_{22})\in\mathbb{R}^4_+$ satisfy this set
of inequalities, then we must have
\begin{equation}
r_{11}+r_{21}=\frac{1}{2}\log\frac{\sigma^2_X}{D^*_3(R,R)}
\label{c06}
\end{equation}
and
\begin{equation}
r_{12}=r_{22}=\frac{1}{2}\log\left(\frac{2D^*_3(R,R)\sigma^2_X}{2D^*_3(R,R)\sigma^2_X-\sigma^2_X\sigma^2_N+D^*_3(R,R)\sigma^2_N}\right).\label{c07}
\end{equation}
From equation (\ref{c07}), it is easy to get that
\begin{equation}
\zeta|_{D_3=D^*_3(R,R)}=D^*_3(R,R)-\left(\frac{1}{\sigma^2_X}+\frac{2}{\sigma^2_N}\right)^{-1}.
\end{equation}
Equations (\ref{c04}) and (\ref{c06}) implie that
\begin{equation}
\lambda(D_1,D_2,D_3,r_{21},r_{22})=0,
\end{equation}
which further implies that $\zeta|_{D_3=D^*_3(R,R)}\leq
d_1+d_2-d_X$, i.e.,
\begin{equation}
D_1+D_2\geq\zeta|_{D_3=D^*_3(R,R)}+\sigma^2_X+\left(\frac{1}{\sigma^2_X}+\frac{2}{\sigma^2_N}\right)^{-1}=D^*_3(R,R)+\sigma^2_X.\label{c10}
\end{equation}
By (\ref{c01}) and (\ref{c03}), we have
\begin{eqnarray*}
r_{i1}&\leq&\min\left(\frac{1}{2}\log\left(1+\frac{\sigma^2_X-\sigma^2_X\exp(-2r_{i2})}{\sigma^2_N}\right),R-r_{i2}\right)\nonumber\\
&=&\min\left(\frac{1}{2}\log\frac{\sigma^2_X+D^*_3(R,R)}{2D^*_3(R,R)},\frac{1}{4}\log\frac{\sigma^2_X}{D^*_3(R,R)}\right)\nonumber\\
&=&\frac{1}{4}\log\frac{\sigma^2_X}{D^*_3(R,R)},\quad i=1,2,
\end{eqnarray*}
which, together with (\ref{c06}), implies
\begin{equation*}
r_{11}=r_{21}=\frac{1}{4}\log\frac{\sigma^2_X}{D^*_3(R,R)}.
\end{equation*}
Thus by (\ref{c02}), we obtain
\begin{equation}
D_i\geq\sigma^2_X\exp(-2r_{i1})=\sigma_X\sqrt{D^*_3(R,R)},\quad
i=1,2. \label{c11}
\end{equation}
Combining (\ref{c10}) and (\ref{c11}) yields
\begin{equation}
\mathcal{D}_{12}^{out}(R)=\left\{(D_1,D_2):D_1+D_2\geq
D^*_3(R,R)+\sigma^2_X,
D_i\geq\sigma_X\sqrt{D^*_3(R,R)},i=1,2\right\}.
\end{equation}

The main technical difficulty of computing
$\mathcal{D}^{in}_{12}(R)$ lies in the convexification operation.
Fortunately, the following lemma significantly reduces the
computational complexity.
\begin{lemma}
For any $\lambda_1,\lambda_2\in (0,1)$ with
$\lambda_1+\lambda_2=1$, and $(R'_1,R'_2), (R''_1,R''_2)$ with
$\lambda_1 (R'_1,R'_2)+\lambda_2 (R''_1,R''_2)=(R,R)$, we have
$\lambda_1
D^*_3(R'_1,R'_2)+\lambda_2D^*_3(R''_1,R''_2)>D^*_3(R,R)$ if
$R'_1+R'_2\neq R''_1+R''_2$.
\end{lemma}
\begin{proof}
See Appendix III.
\end{proof}

This lemma implies that it is impossible to achieve $D^*_3(R,R)$
by timesharing two distributed source coding schemes, one with the
sum-rate higher than $2R$ and the other with the sum-rate lower
than $2R$. Therefore, we have
\begin{eqnarray*}
\mathcal{D}^{in}_{12}(R)=\left\{(D_1,D_2): (R,R,D_1,D_2)\in
conv\left(\bigcup\left(\mathcal{A}_1
\cup\mathcal{A}_2\right)\right)\right\},
\end{eqnarray*}
where
\begin{eqnarray*}
\mathcal{A}_1\triangleq\left\{(R_1,R_2,D_1,D_2):
\frac{1}{D_i}\leq\frac{1}{\sigma_X^2}+\frac{1}{\sigma^2_{N_i}+\sigma^2_{T_{i1}}}, i=1,2,\right.\\
\left.
R_1=\frac{1}{2}\log\frac{\sigma^2_{U_1}(\sigma^2_{W_{1}}\sigma^2_{W_{2}}-\sigma^4_X)}{\sigma^2_{T_{12}}(\sigma^2_{U_{1}}\sigma^2_{W_{2}}-\sigma^4_X)},
 R_1+R_2=2R\right\},
\end{eqnarray*}
\begin{eqnarray*}
\mathcal{A}_2\triangleq\left\{(R_1,R_2,D_1,D_2):
\frac{1}{D_i}\leq\frac{1}{\sigma_X^2}+\frac{1}{\sigma^2_{N_i}+\sigma^2_{T_{i1}}}, i=1,2,\right.\\
\left.
R_1=\frac{1}{2}\log\frac{\sigma^2_{U_1}(\sigma^2_{W_{1}}\sigma^2_{U_{2}}-\sigma^4_X)}{\sigma^2_{T_{12}}(\sigma^2_{U_{1}}\sigma^2_{U_{2}}-\sigma^4_X)},
R_1+R_2=2R\right\}.
\end{eqnarray*}
and $\bigcup$ is taken over all
$(\sigma^2_{T_{11}},\sigma^2_{T_{12}},\sigma^2_{T_{21}},\sigma^2_{T_{21}})$
such that $\sigma^2_{T_{11}}\geq\sigma^2_{T_{12}}$,
$\sigma^2_{T_{21}}\geq\sigma^2_{T_{21}}$ and
\begin{eqnarray}
2R&=&\frac{1}{2}\log\frac{\sigma^2_{U_{1}}\sigma^2_{U_{2}}(\sigma^2_{W_{1}}\sigma^2_{W_{2}}-\sigma^4_X)}{\sigma^2_{T_{12}}\sigma^2_{T_{22}}(\sigma^2_{U_{1}}\sigma^2_{U_{2}}-\sigma^4_X)},\label{s1}\\
\frac{1}{D^*_3(R,R)}&=&\frac{1}{\sigma_X^2}+\frac{1}{\sigma^2_{N}+\sigma^2_{T_{12}}}+\frac{1}{\sigma^2_{N}+\sigma^2_{T_{22}}}.\label{s2}
\end{eqnarray}
Let
\begin{eqnarray*}
R^*&=&\frac{1}{2}\log\frac{4D^*_3(R,R)\sigma^4_X}{[\sigma^2_X+D^*_3(R,R)][2D^*_3(R,R)\sigma^2_X-\sigma^2_X\sigma^2_N+D^*_3(R,R)\sigma^2_N]},\\
\varphi(x)&=&\frac{\left[2D^*_3(R,R)\sigma^2_X-\sigma^2_X\sigma^2_N+D^*_3(R,R)\sigma^2_N\right]\exp\left[2(2R-x)\right]-2D^*_3(R,R)\sigma^2_X}{\sigma^2_X-D^*_3(R,R)}.
\end{eqnarray*}
Note (\ref{s1}) and (\ref{s2}) imply that
\begin{eqnarray*}
\sigma^2_{T_{12}}=\sigma^2_{T_{22}}=\frac{2\sigma^2_XD^*_3(R,R)-\sigma^2_X\sigma^2_N+\sigma^2_ND^*_3(R,R)}{\sigma^2_X-D^*_3(R,R)}
\end{eqnarray*}
and $\sigma^2_{T_{12}}\sigma^2_{T_{22}}=\infty$ (i.e.,
$\sigma^2_{T_{12}}=\infty$ or $\sigma^2_{T_{22}}=\infty$).
Therefore,
\begin{eqnarray*}
\mathcal{D}^{in}_{12}(R)=\left\{(D_1,D_2): (R,D_1,D_2)\in
conv\left(\widetilde{\mathcal{A}}_1
\cup\widetilde{\mathcal{A}}_2\right)\right\},
\end{eqnarray*}
where
\begin{eqnarray*}
\widetilde{\mathcal{A}}_1&=&\left\{(\widetilde{R},D_1,D_2):
R^*\leq \widetilde{R}\leq
2R-R^*, D_1\geq\varphi(\widetilde{R}), D_2=\sigma^2_X\right\},\\
\widetilde{\mathcal{A}}_2&=&\left\{(\widetilde{R},D_1,D_2):
R^*\leq \widetilde{R}\leq 2R-R^*, D_1=\sigma^2_X,
D_2\geq\varphi(2R-\widetilde{R})\right\}.
\end{eqnarray*}
Let
$\partial\mathcal{D}^{in}_{12}(R)=\{(D_1,D_2)\in\mathcal{D}^{in}_{12}(R):
(D'_1\leq D_1,D'_2\leq D_2)\Rightarrow (D'_1=D_1,D'_2=D_2),
\forall (D'_1,D'_2)\in\mathcal{D}^{in}_{12}(R) \}$. It is clear
that $\mathcal{D}^{in}_{12}(R)$ is completely characterized by
$\partial\mathcal{D}^{in}_{12}(R)$. Note that
$\widetilde{\mathcal{A}}_1\cup\widetilde{\mathcal{A}}_2$ is a
subset of a 3-dimensional linear space. Thus by
Carath$\acute{\mbox{e}}$odory's fundamental theorem
\cite{Eggleston}, for any
$(D_1,D_2)\in\partial\mathcal{D}^{in}_{12}(R)$, $(R,D_1,D_2)$ can
be expressed as a convex combination of at most 4 points in
$\widetilde{\mathcal{A}}_1\cup\widetilde{\mathcal{A}}_2$. Actually
this can be further simplified. Since $\varphi(x)$ is a convex
function, it implies that for any
$(D_1,D_2)\in\partial\mathcal{D}^{in}_{12}(R)$, $(R,D_1,D_2)$ can
be expressed as a convex combination of a point
$(R',D'_1,D'_2)\in\widetilde{\mathcal{A}}_1$ with
$D'_1=\varphi(R')$ and a point
$(R'',D''_1,D''_2)\in\widetilde{\mathcal{A}}_2$ with
$D''_2=\varphi(2R-R'')$. Now the problem is readily solved by
Lagrangian optimization. Through tedious but straightforward
calculation, $\partial\mathcal{D}^{in}_{12}(R)$ is the curve
$D_1=\psi(D_2)$ given by the following parametric form:
\begin{eqnarray*}
&\left\{\begin{array}{c}
  D_1=\frac{R-R^*}{2R-R^*-\mu}\sigma^2_X+\frac{R-\mu}{2R-R^*-\mu}\varphi(2R-R^*) \\
  \hspace{-0.1in}D_2=\frac{R-R^*}{2R-R^*-\mu}\varphi(2R-\mu)+\frac{R-\mu}{2R-R^*-\mu}\sigma^2_X
\end{array}\right. \mbox{for } R^*\leq\mu\leq R,
\end{eqnarray*}
\begin{eqnarray*}
&\left\{\begin{array}{c}
  \hspace{-0.5in}D_1=\frac{R-\mu}{R^*-\mu}\sigma^2_X+\frac{R^*-R}{R^*-\mu}\varphi(\mu) \\
  D_2=\frac{R-\mu}{R^*-\mu}\varphi(2R-R^*)+\frac{R^*-R}{R^*-\mu}\sigma^2_X
\end{array}\right. \mbox{for } R<\mu\leq 2R-R^*.
\end{eqnarray*}
Hence we have
\begin{eqnarray*}
\mathcal{D}^{in}_{12}(R)=\left\{(D_1,D_2):D_1\geq\psi(D_2),
D_i\geq\frac{2\sqrt{D^*_3(R,R)}\sigma^2_X}{\sigma_X+\sqrt{D^*_3(R,R)}},
i=1,2\right\}.
\end{eqnarray*}

As we can see in Fig. 4, $\mathcal{D}^{out}_{12}(R)$ is strictly
bigger than $\mathcal{D}^{in}_{12}(R)$.

\begin{figure}[hbt]
\centering
\includegraphics[scale=0.5]{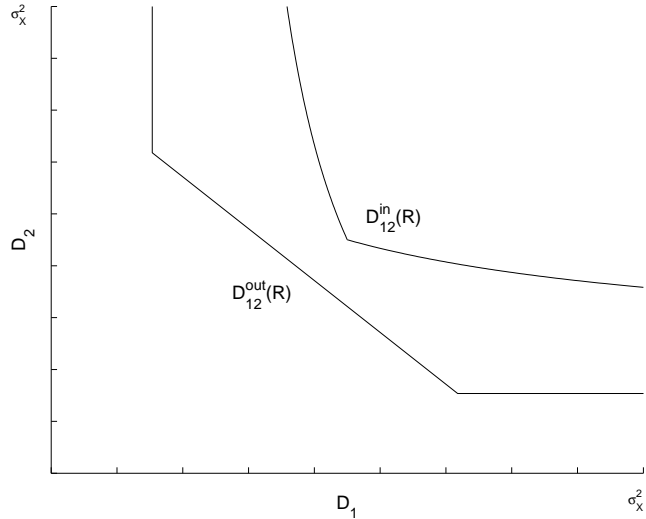}
\caption{Comparison of $\mathcal{D}^{in}_{12}(R)$ and $\mathcal{D}^{out}_{12}(R)$}
\end{figure}

\subsection{An Extreme Case}

\begin{theorem}\label{e1} Let
$\mathcal{Q}_e=\{(R_1,D_1,D_2,D_3):(R_1,\infty,D_1,D_2,D_3)\in\mathcal{Q}\}$.
We have $(R_1,D_1,D_2,D_3)\in\mathcal{Q}_e$ if and only if
$D_2\geq D_{2,\min}$ and
\begin{eqnarray*}
R_1\geq\left\{
\begin{array}{lc}
  \frac{1}{2}\log\frac{\sigma_X^4\sigma_{N_2}^4}{\left(\sigma_{N_2}^2+D_1\right)\left(D_3\sigma_X^2\sigma_{N_1}^2+D_3\sigma_X^2\sigma_{N_2}^2+D_3\sigma_{N_1}^2\sigma_{N_2}^2-\sigma_X^2\sigma_{N_1}^2\sigma_{N_2}^2\right)}, & D_3\leq\frac{D_1\sigma^2_{N_2}}{D_1+\sigma^2_{N_2}} \\
  R(D_1,\sigma^2_{N_1}),
  & D_3>\frac{D_1\sigma^2_{N_2}}{D_1+\sigma^2_{N_2}}.
\end{array}
 \right.
\end{eqnarray*}
\end{theorem}
\begin{proof}
Since $R_2=\infty$, we can assume that $\{Y_2(t)\}_{t=1}^\infty$
is directly present at decoder 2 and decoder 3. Hence any $D_2\geq
D_{2,\min}$ is achievable. Now only $(R_1,D_1,D_3)$ remain to be
characterized. The achievability part follows directly by
evaluating $\mathcal{Q}_{in}$ with
$\sigma^2_{T_{21}}=\sigma^2_{T_{22}}=0$. For the converse, it is
clear that $R_1\geq R(D_1,\sigma^2_{N_1})$, which resolves the
case $D_3>{D_1\sigma^2_{N_2}}/{(D_1+\sigma^2_{N_2})}$. For the
case $D_3\leq {D_1\sigma^2_{N_2}}/{(D_1+\sigma^2_{N_2})}$, the
details are left to Appendix IV.
\end{proof}
Remark: The converse can not be reduced from $\mathcal{Q}_{out}$,
which shows that our outer bound in not tight.

Theorem \ref{e1} implies that
\begin{equation*}
D^*_3(R_1,\infty)=\frac{\sigma_X^4\sigma_{N_2}^4\exp\left(-2R_1\right)+\sigma_X^2\sigma_{N_1}^2\sigma_{N_2}^2\left(\sigma_X^2+\sigma_{N_2}^2\right)}
{\left(\sigma_X^2+\sigma_{N_2}^2\right)\left(\sigma_X^2\sigma_{N_1}^2+\sigma_X^2\sigma_{N_2}^2+\sigma_{N_1}^2\sigma_{N_2}^2\right)},
\end{equation*}
and
$\min\{D_1:(R_1,\infty,D_1,D_2,D^*_3(R_1,\infty))\in\mathcal{Q}\}=\sigma^2_X$
for $D_2\geq D_{2,\min}$. That is to say, for this extreme case,
if decoder 3 achieves the minimum $D_3$ for a given $R_1$, then it
is impossible for decoder 1 to make a nontrivial estimation of
$\{X(t)\}_{t=1}^\infty$.

\subsection{Noisy Multiple Description for the Gaussian Case}

Now consider the case when both encoder 1 and encoder 2 can
observe $\{Y_1(t)\}_{t=1}^\infty$ and $\{Y_2(t)\}_{t=1}^\infty$
simultaneously. Clearly, the rate distortion region of this
problem (which we denote by $\mathcal{Q}'$) is an outer bound of
$\mathcal{Q}$.

If we assume encoder 1 and encoder 2 can only observe
$\{\mathbb{E}(X(t)|Y_1(t),Y_2(t))\}_{t=1}^\infty$, and let
$\mathcal{Q}''$ be the rate distortion region for this case, then
clearly we have $\mathcal{Q}''\subseteq\mathcal{Q}'$ since
$\{\mathbb{E}(X(t)|Y_1(t),Y_2(t))\}_{t=1}^\infty$ can be computed
from $\{Y_1(t)\}_{t=1}^\infty$ and $\{Y_2(t)\}_{t=1}^\infty$.

\begin{theorem}\label{n1}
$\mathcal{Q}'=\mathcal{Q}''=\mathcal{T}$, where
\begin{eqnarray*}
\mathcal{T}=\left\{(R_1,R_2,D_1,D_2,D_3): R_1+R_2\geq
\frac{1}{2}\log\frac{d_X}{d_3}+\gamma(d_1,d_2,d_3),
R_i\geq\frac{1}{2}\log\frac{d_X}{d_i}, i=1,2\right\} ,
\end{eqnarray*}
and
\begin{eqnarray*}
&\gamma(d_1,d_2,d_3)=\left\{\begin{array}{ll}
  0, \hspace{2.8in} d_3\leq d_1+d_2-d_X\\
  \frac{1}{2}\log\frac{d_Xd_3}{d_1d_2}, \hspace{2.21in} d_3\geq \left(\frac{1}{d_1}+\frac{1}{d_2}-\frac{1}{d_X}\right)^{-1}\\
  \frac{1}{2}\log\frac{(d_X-d_3)^2}
{(d_X-d_3)^2-[\sqrt{(d_X-d_1)(d_X-d_2)}-\sqrt{(d_1-d_3)(d_2-d_3)}]^2},
\mbox{\quad otherwise}.
\end{array}\right.&\\
\end{eqnarray*}
\end{theorem}
\begin{proof}
As defined before, let $S(t)=\mathbb{E}(X(t)|Y_1(t),Y_2(t))$ and
$\theta(t)=X(t)-S(t)$, $t=1,2,\cdots$. We have
$\mathbb{E}\theta^2(t)=D_{3,\min}$ and
$\mathbb{E}S^2(t)=\sigma^2_X-D_{3,\min}=d_X$.

Now we view $\{S(t)\}_{t=1}^{\infty}$ as the source, and let
$\mathcal{Q}_S$ be the multiple description rate-distortion region
for $\{S(t)\}_{t=1}^{\infty}$. It was proved by Ozarow
\cite{Ozarow} that $(R_1,R_2,D_1,D_2,D_3)\in\mathcal{Q}_S$ if and
only if
\begin{eqnarray*}
R_i&\geq&\frac{1}{2}\log\frac{d_X}{D_i},\quad i=1,2,\\
R_3&\geq&\frac{1}{2}\log\frac{d_X}{D_3}+\gamma_S(D_1,D_2,D_3),
\end{eqnarray*}
where
\begin{eqnarray*}
&\gamma_S(D_1,D_2,D_3)=\left\{\begin{array}{ll}
  0, \hspace{2.8in} D_3\leq D_1+D_2-d_X\\
  \frac{1}{2}\log\frac{d_XD_3}{D_1D_2}, \hspace{2.21in} D_3\geq \left(\frac{1}{D_1}+\frac{1}{D_2}-\frac{1}{d_X}\right)^{-1}\\
  \frac{1}{2}\log\frac{(d_X-D_3)^2}
{(d_X-D_3)^2-[\sqrt{(d_X-D_1)(d_X-D_2)}-\sqrt{(D_1-D_3)(D_2-D_3)}]^2},
\mbox{\quad otherwise}.
\end{array}\right.&\\
\end{eqnarray*}

Since $\{\theta(t)\}_{t=1}^n$ is independent of
$\{Y_1(t),Y_2(t)\}_{t=1}^n$ and thus is independent of $\{\hat
X_1(t),\hat X_2(t),\hat X_3(t)\}_{t=1}^n$, we have
\begin{eqnarray*}
\frac{1}{n}\mathbb{E}\sum\limits_{t=1}^n(X(t)-\hat
X_i(t))^2&=&\frac{1}{n}\sum\limits_{t=1}^n\mathbb{E}(S(t)+\theta(t)-\hat
X_i(t))^2\\
&=&\frac{1}{n}\sum\limits_{t=1}^n\left[\mathbb{E}\theta^2(t)+\mathbb{E}(S(t)-\hat
X_i(t))^2\right]\\
&=&D_{3,\min}+\frac{1}{n}\sum\limits_{t=1}^n\mathbb{E}(S(t)-\hat
X_i(t))^2,\quad i=1,2,3.
\end{eqnarray*}
Hence $(R_1,R_2,D_1,D_2,D_3)\in\mathcal{Q}'$(or $\mathcal{Q}''$)
if and only if
$(R_1,R_2,D_1-D_{3,\min},D_2-D_{3,\min},D_3-D_{3,\min})\in\mathcal{Q}_S$
(i.e., $(R_1,R_2,d_1,d_2,d_3)\in\mathcal{Q}_S$). The proof is
complete.
\end{proof}

Remark: Theorem \ref{n1} is still true when $N_1(t)$ and $N_2(t)$
are correlated (with correlation coefficient $\rho_N$), now
\begin{eqnarray*}
D_{3,\min}=\left(\frac{1}{\sigma^2_X}+\frac{\sigma^2_{N_1}+\sigma^2_{N_2}-2\rho_N\sigma_{N_1}\sigma_{N_2}}{(1-\rho_N^2)\sigma^2_{N_1}\sigma^2_{N_2}}\right)^{-1}.
\end{eqnarray*}
Note: $D_{3,\min}=(1/\sigma^2_X+1/\sigma^2_N)^{-1}$ if
$\sigma^2_{N_1}=\sigma^2_{N_2}=\sigma^2_N$ and $\rho_N=1$.

\section{Conclusion}
We proposed  a robust distributed source coding scheme which
flexibly trades off between system robustness and compression
efficiency.  The achievable rate distortion region of this scheme
was analyzed in detail for the Gaussian case. But a complete
characterization of the rate distortion region $\mathcal{Q}$, even
for the Gaussian case, remains open. We believe that the following
problem deserves special attention. For the Gaussian case,
$\mathcal{Q}|_{\sigma^2_{N_1}=\sigma^2_{N_2}=0}$ is the rate
distortion region of the multiple description problem, which has
been completely characterized in \cite{Ozarow}. The question is
whether $\mathcal{Q}$ converges to
$\mathcal{Q}|_{\sigma^2_{N_1}=\sigma^2_{N_2}=0}$ as
$\sigma^2_{N_1}$ and $\sigma^2_{N_2}$ go to zero. A solution to
this problem will have many interesting implications and can
significantly deepen our understanding of the multiple description
problem and the distributed source coding problem.

\appendices

\section{Proof of Theorem 1}
The proof of Theorem 1 employs techniques which have already been
established in the literature, especially in
\cite{Han}\cite{Heegard}\cite{BergerYeung}\cite{AlshwedeKorner}.
Hence we only give a sketch here.

For each $U_1, U_2, W_1$ and $W_2$ satisfying Property (i) and
(iii), we prove the admissibility of the rate tuple $(R_1,R_2)$,
where
\begin{eqnarray*}
R_1&=&I(Y_1;U_1)+I(Y_1;W_1|U_1,U_2,W_2),\\
R_2&=&I(Y_2;U_2)+I(Y_2;W_2|U_1,U_2).
\end{eqnarray*}
Then by symmetry, the rate tuple $(R'_1,R'_2)$ with
\begin{eqnarray*}
R'_1&=&I(Y_1;U_1)+I(Y_1;W_1|U_1,U_2),\\
R'_2&=&I(Y_2;U_2)+I(Y_2;W_2|U_1,U_2,W_1)
\end{eqnarray*}
is also admissible. It's easy to check that
\begin{eqnarray*}
&&I(Y_1;W_1|U_1,U_2,W_2)+I(Y_2;W_2|U_1,U_2)\\&=&I(Y_1;W_1|U_1,U_2)+I(Y_2;W_2|U_1,U_2,W_1)\\
&=&I(Y_1,Y_2;W_1,W_2|U_1,U_2),
\end{eqnarray*}
now Theorem 1 follows by timesharing $(R_1,R_2)$ and
$(R'_1,R'_2)$.

It was established in \cite{Heegard} that for any positive
$\epsilon$ and sufficiently large $n$ with
\begin{equation*}
|\mathcal{C}^{(n)}_1|\leq
\exp(n(I(Y_1;U_1)+I(Y_1;W_1|U_1,U_2,W_2)+\epsilon)),
\end{equation*}
decoder 1 and decoder 3 can recover $u^n_1$ and construct $\hat
x^n_1$ with $\hat x_1(t)=f_1(u_1(t)), t=1,2,\cdots,n,$ such that
\begin{equation*}
\frac{1}{n}E\sum\limits_{t=1}^n d(X(t), \hat X_1(t))<D_1+\epsilon,
\end{equation*}
and provided $u_2^n$ and $w_2^n$ are available to decoder 3, it
can further recover $w_1^n$ and use $w^n_1,w^n_2,u^n_1,u^n_2$ to
construct $\hat x^n_3$ with $\hat
x_3(t)=f_3\left(w_1(t),w_2(t),u_1(t),u_2(t)\right),
t=1,2,\cdots,n,$ such that the average distortion is less than or
equal to $D_3+\epsilon$.

Again by \cite{Heegard}, with
\begin{equation*}
|\mathcal{C}^{(n)}_2|\leq
\exp(n(I(Y_2;W_2)+I(Y_2;U_2|W_1,W_2)+\epsilon)),
\end{equation*}
decoder 2 and decoder 3 can recover $u^n_2$ and construct $\hat
x^n_2$ with $\hat x_2(t)=f_2(u_2(t)), t=1,2,\cdots,n,$ such that
\begin{equation*}
\frac{1}{n}E\sum\limits_{t=1}^n d(X(t), \hat X_2(t))<D_2+\epsilon,
\end{equation*}
and provided $u_1^n$ are available to decoder 3, it can further
recover $w_2^n$.

In summary, decoder $i$ recovers $u^n_i$ $(i=1,2)$, and decoder 3
recovers $u^n_1,u^n_2,w^n_1,w^n_2$ with the decoding order
$(u^n_1, u^n_2)\rightarrow w^n_2\rightarrow w^n_1$.

Thus we have established the admissibility of the rate tuple
$(R_1,R_2)$ and completed the proof.

\section{Derivation of the Outer bound}
Let
$r_{i1}=I(X^n;f^{(n)}_{E,i}(Y^n_i))/n,r_{i2}=I(Y_i^n;f^{(n)}_{E,i}(Y^n_i)|X^n)/n$,
$i=1,2$. The following lemmas were proved in \cite{TseCEO} with
the method developed by Oohama \cite{OohamaCEO}.
\begin{lemma}\label{lemmaApp2}
\begin{eqnarray*}
\frac{1}{\sigma_X^2}\exp\left(2r_{i1}\right)&\leq&\frac{1}{\sigma_X^2}+\frac{1-\exp\left(2r_{i2}\right)}{\sigma_{N_i}^2}\quad i=1,2,\\
\frac{1}{\sigma_X^2}\exp\left(\frac{2}{n}I\left(X^n;f^{(n)}_{E,1}(Y^n_1)f^{(n)}_{E,2}(Y^n_2)\right)\right)&\leq&\frac{1}{\sigma_X^2}+\frac{1-\exp\left(-2r_{12}\right)}{\sigma_{N_i}^2}+\frac{1-\exp(-2r_{22})}{\sigma_{N_2}^2}.
\end{eqnarray*}
\end{lemma}

Now we are ready to derive the outer bound.

\begin{proof}
By data processing inequality, we have
\begin{eqnarray}
I\left(X^n;f^{(n)}_{E,i}(Y^n_i)\right)\geq I\left(X^n;\hat
X^n_i\right)\geq\frac{n}{2}\log\frac{\sigma_X^2}{D_i},\quad
i=1,2\label{10}\\I\left(X^n;f^{(n)}_{E,1}(Y^n_1)f^{(n)}_{E,2}(Y^n_2)\right)\geq
I\left(X^n;\hat
X^n_3\right)\geq\frac{n}{2}\log\frac{\sigma_X^2}{D_3}.\label{11}
\end{eqnarray}
It follows from (\ref{10}), (\ref{11}) and Lemma \ref{lemmaApp2}
that
\begin{eqnarray*}
\frac{1}{D_i}&\leq&\frac{\exp(2r_{i1})}{\sigma^2_X},\quad i=1,2,\\
\frac{1}{D_3}&\leq&\frac{1}{\sigma^2_X}+\frac{1-\exp(-2r_{12})}{\sigma^2_{N_1}}+\frac{1-\exp(-2r_{12})}{\sigma^2_{N_1}}.
\end{eqnarray*}

Since $X^n\rightarrow Y^n_i\rightarrow f^{(n)}_{E,i}(Y^n_i)$,
$i=1,2$, we have
\begin{eqnarray*}
R_i&\geq&\frac{1}{n} I\left(Y^n_i;f^{(n)}_{E,i}(Y^n_i)\right)\geq
\frac{1}{n}I\left(X^n,Y^n_i;f^{(n)}_{E,i}(Y^n_i)\right)\\&=&\frac{1}{n}I\left(X^n;f^{(n)}_{E,i}(Y^n_i)\right)+\frac{1}{n}I\left(\left.Y_i^n;f^{(n)}_{E,i}(Y^n_i)\right|X^n\right)\\
&=&r_{i1}+r_{i2}\quad i=1,2.
\end{eqnarray*}

Now we proceed to derive a lower bound on $r_{11}+r_{21}$,
\begin{eqnarray}
&&n(r_{11}+r_{21})\nonumber\\
&=&I\left(X^n;f^{(n)}_{E,1}(Y^n_1)\right)+I\left(X^n;f^{(n)}_{E,2}(Y^n_2)\right)\nonumber\\
&\stackrel{(a)}{=}&I\left(X^n;f^{(n)}_{E,1}(Y^n_1),f^{(n)}_{E,2}(Y^n_2)\right)+I\left(f^{(n)}_{E,1}(Y^n_1);f^{(n)}_{E,2}(Y^n_2)\right)
-I\left(\left.f^{(n)}_{E,1}(Y^n_1);f^{(n)}_{E,2}(Y^n_2)\right|X^n\right)\nonumber\\
&\stackrel{(b)}{=}&I\left(X^n;f^{(n)}_{E,1}(Y^n_1),f^{(n)}_{E,2}(Y^n_2)\right)+I\left(f^{(n)}_{E,1}(Y^n_1);f^{(n)}_{E,2}(Y^n_2)\right)\label{9}
\end{eqnarray}
where (a) follows from the identity
\begin{equation}
I\left(A;BC\right)=I\left(A;B\right)+I\left(A;C\right)+I\left(B;C|A\right)-I\left(B;C\right).\label{12}
\end{equation}
and (b) is because $f^{(n)}_{E,1}(Y^n_1)\rightarrow X^n\rightarrow
f^{(n)}_{E,2}(Y^n_2)$. Now applying data processing inequality, we
have
\begin{eqnarray}
n\left(r_{11}+r_{21}\right)&\geq& I\left(X^n;\hat
X^n_3\right)+I\left(f^{(n)}_{E,1}(Y^n_1);f^{(n)}_{E,2}(Y^n_2)\right)\nonumber\\
&&\geq\frac{1}{2}\log\frac{\sigma^2_X}{D_3}+I\left(f^{(n)}_{E,1}(Y^n_1);f^{(n)}_{E,2}(Y^n_2)\right).\label{23}
\end{eqnarray}

To lower-bound
$I\left(f^{(n)}_{E,1}(Y^n_1);f^{(n)}_{E,2}(Y^n_2)\right)$, we
introduce an auxiliary random vector $Z^n$ such that
$Z(t)=S(t)+M(t),t=1,2,\cdots,n$, where the $M(t)'s$ are i.i.d
zero-mean Gaussian random variables with variance $\sigma^2_M$
(which will be optimized later). We assume that $M^n$ is
independent of $(X^n, Y^n_1, Y^n_2)$. Since $\theta^n$ is
indepedent of $Y^n_1, Y^n_2$ and thus independent of $\hat X^n_1,
\hat X^n_2$, we have
\begin{eqnarray*}
D_i&\geq&\frac{1}{n}\sum\limits_{t=1}^n{\mathbb{E}\left(X(t)-\hat
X_i(t)\right)^2}\\&=&\frac{1}{n}\sum\limits_{t=1}^n{\mathbb{E}(S(t)+\theta(t)-\hat
X_i(t))^2}\\&=&\frac{1}{n}\sum\limits_{t=1}^n{\mathbb{E}\left(S(t)-\hat
X_i(t)\right)^2}+D_{3,\min},
\end{eqnarray*}
i.e.,
\begin{equation*}
\frac{1}{n}\sum{\mathbb{E}\left(S(t)-\hat X_i(t)\right)^2}\leq
D_i-D_{3,\min}=d_i,\quad i=1,2.
\end{equation*}
Since
\begin{eqnarray*}
\frac{1}{n}\sum{\mathbb{E}\left(Z(t)-\hat
X_i(t)\right)^2}&=&\frac{1}{n}\sum{\mathbb{E}\left(S(t)-\hat
X_i(t)\right)^2}+\frac{1}{n}\sum{\mathbb{E}M^2(t)}\\&\leq&
d_i+\sigma^2_M, \quad i=1,2,
\end{eqnarray*}
by rate distortion theory,
\begin{equation*}
I\left(Z^n;\hat
X^n_i\right)\geq\frac{n}{2}\log\left(\frac{d_X+\sigma^2_M}{d_i+\sigma^2_M}\right),
\quad i=1,2.
\end{equation*}
Now applying the identity (\ref{12}) to
$I\left(f^{(n)}_{E,1}(Y^n_1);f^{(n)}_{E,2}(Y^n_2)\right)$, we get
\begin{eqnarray}
&&I\left(f^{(n)}_{E,1}(Y^n_1);f^{(n)}_{E,2}(Y^n_2)\right)\nonumber\\
&=&I\left(Z^n;f^{(n)}_{E,1}(Y^n_1)\right)+I\left(Z^n;f^{(n)}_{E,2}(Y^n_2)\right)+I\left(\left.f^{(n)}_{E,1}(Y^n_1);f^{(n)}_{E,2}(Y^n_2)\right|Z^n\right)-I\left(Z^n;f^{(n)}_{E,1}(Y^n_1),f^{(n)}_{E,2}(Y^n_2)\right)\nonumber\\
&\geq&I\left(Z^n;\hat X^n_1\right)+I\left(Z^n;\hat
X^n_2\right)-I\left(Z^n;f^{(n)}_{E,1}(Y^n_1),f^{(n)}_{E,2}(Y^n_2)\right)\nonumber\\
&\geq&\frac{n}{2}\log\left[\left(\frac{d_X+\sigma^2_M}{d_1+\sigma^2_M}\right)\left(\frac{d_X+\sigma^2_M}{d_2+\sigma^2_M}\right)\right]-I\left(Z^n;f^{(n)}_{E,1}(Y^n_1),f^{(n)}_{E,2}(Y^n_2)\right).\label{13}
\end{eqnarray}

We upper-bound
$I\left(Z^n;f^{(n)}_{E,1}(Y^n_1),f^{(n)}_{E,2}(Y^n_2)\right)$ as
follows:
\begin{eqnarray}
&&I\left(Z^n;f^{(n)}_{E,1}(Y^n_1),f^{(n)}_{E,2}(Y^n_2)\right)\nonumber\\
&=&h(Z^n)-h\left(Z^n\left|f^{(n)}_{E,1}(Y^n_1),f^{(n)}_{E,2}(Y^n_2)\right.\right)\nonumber\\
&=&\frac{n}{2}\log\left[2\pi
e\left(d_X+\sigma^2_M\right)\right]-h\left(S^n+M^n|f^{(n)}_{E,1}(Y^n_1),f^{(n)}_{E,2}(Y^n_2)\right)\nonumber\\
&\stackrel{(c)}{\leq}&\frac{n}{2}\log\left[2\pi
e\left(d_X+\sigma^2_M\right)\right]-\frac{n}{2}\log\left\{\exp\left[\frac{2}{n}h(S^n|f^{(n)}_{E,1}(Y^n_1),f^{(n)}_{E,2}(Y^n_2))\right]+2\pi
e\sigma^2_M\right\},\label{14}
\end{eqnarray}
where (c) follows from the conditional version of entropy power
inequality \cite{Blachman}. Since
\begin{eqnarray*}
&&h\left(S^n\left|f^{(n)}_{E,1}(Y^n_1),f^{(n)}_{E,2}(Y^n_2)\right.\right)\\
&=&h\left(S^n\left|X^n,f^{(n)}_{E,1}(Y^n_1),f^{(n)}_{E,2}(Y^n_2)\right.\right)+I\left(X^n;\hat X^n\left|f^{(n)}_{E,1}(Y^n_1),f^{(n)}_{E,2}(Y^n_2)\right.\right)\\
&=&h\left(S^n\left|X^n,f^{(n)}_{E,1}(Y^n_1),f^{(n)}_{E,2}(Y^n_2)\right.\right)+I\left(X^n;S^n,f^{(n)}_{E,1}(Y^n_1),f^{(n)}_{E,2}(Y^n_2)\right)-I\left(X^n;f^{(n)}_{E,1}(Y^n_1),f^{(n)}_{E,2}(Y^n_2)\right)\\
&=&h\left(S^n\left|X^n,f^{(n)}_{E,1}(Y^n_1),f^{(n)}_{E,2}(Y^n_2)\right.\right)+I\left(X^n;S^n\right)-I\left(X^n;f^{(n)}_{E,1}(Y^n_1),f^{(n)}_{E,2}(Y^n_2)\right),
\end{eqnarray*}
where the last equality follows from $X^n\rightarrow
S^n\rightarrow\left(f^{(n)}_{E,1}(Y^n_1),f^{(n)}_{E,2}(Y^n_2)\right)$,
we have
\begin{eqnarray}
&&\exp\left(\frac{2}{n}h\left(S^n\left|f^{(n)}_{E,1}(Y^n_1),f^{(n)}_{E,2}(Y^n_2)\right.\right)\right)\nonumber\\
&=&\exp\left(h\left(S^n\left|X^n,f^{(n)}_{E,1}(Y^n_1),f^{(n)}_{E,2}(Y^n_2)\right.\right)\right)\exp\left(I\left(X^n;S^n\right)\right)\exp\left(-I\left(X^n;f^{(n)}_{E,1}(Y^n_1),f^{(n)}_{E,2}(Y^n_2)\right)\right)\nonumber\\
&=&\frac{\sigma^2_X}{D_{3,\min}}\exp\left(h(S^n|X^n,f^{(n)}_{E,1}(Y^n_1),f^{(n)}_{E,2}(Y^n_2))\right)\exp\left(-I(X^n;f^{(n)}_{E,1}(Y^n_1),f^{(n)}_{E,2}(Y^n_2))\right).\label{15}
\end{eqnarray}
Now we shall derive a lower bound on
$\exp\left(\frac{2}{n}h\left(S^n\left|X^n,
f^{(n)}_{E,1}(Y^n_1),f^{(n)}_{E,2}(Y^n_2)\right.\right)\right)$.
Since conditioned on
$\left(X^n,f^{(n)}_{E,1}(Y^n_1),f^{(n)}_{E,2}(Y^n_2)\right)$,
$Y^n_1$ and $Y^n_2$ are independent, by the conditional version of
entropy power inequality \cite{Blachman}, we have
\begin{eqnarray}
&&\exp\left(\frac{2}{n}h\left(S^n\left|X^n, f^{(n)}_{E,1}(Y^n_1),f^{(n)}_{E,2}(Y^n_2)\right.\right)\right)\nonumber\\
&\geq&\sum\limits_{i=1}^2\exp\left(\frac{2}{n}h\left(\left.\frac{D_{3,\min}}{\sigma^2_{N_i}}Y^n_i\right|X^n, f^{(n)}_{E,1}(Y^n_1),f^{(n)}_{E,2}(Y^n_2)\right)\right)\nonumber\\
&=&D^2_{3,\min}\sum\limits_{i=1}^2\frac{1}{\sigma^4_{N_i}}\exp\left(\frac{2}{n}h\left(Y^n_i\left|X^n, f^{(n)}_{E,1}(Y^n_1),f^{(n)}_{E,2}(Y^n_2)\right.\right)\right)\nonumber\\
&=&D^2_{3,\min}\sum\limits_{i=1}^2\frac{1}{\sigma^4_{N_i}}\exp\left(\frac{2}{n}h(Y^n_i|X^n)-\frac{2}{n}I\left(\left.Y^n_i; f^{(n)}_{E,i}(Y^n_i)\right|X^n\right)\right)\nonumber\\
&=&2\pi
eD^2_{3,\min}\sum\limits_{i=1}^2\frac{\exp\left(-2r_{2i}\right)}{\sigma^2_{N_i}}.\label{16}
\end{eqnarray}
Thus by (\ref{15}) and (\ref{16}),
\begin{eqnarray}
&&\exp\left(\frac{2}{n}h\left(S^n\left|f^{(n)}_{E,1}(Y^n_1),f^{(n)}_{E,2}(Y^n_2)\right.\right)\right)\nonumber\\
&\geq& 2\pi
e\sigma^2_XD_{3,\min}\left(\sum\limits_{i=1}^2\frac{\exp\left(-2r_{2i}\right)}{\sigma^2_{N_i}}\right)\exp\left(-\frac{2}{n}I\left(X^n;f^{(n)}_{E,1}(Y^n_1),f^{(n)}_{E,2}(Y^n_2)\right)\right).\label{18}
\end{eqnarray}
Combining (\ref{14}) and (\ref{18}) yields that
\begin{eqnarray*}
&&I\left(Z^n;f^{(n)}_{E,1}(Y^n_1),f^{(n)}_{E,2}(Y^n_2)\right)\leq\frac{n}{2}\log(d_X+\sigma^2_M))\\
&-&\frac{n}{2}\log\left(\sigma^2_XD_{3,\min}\left(\sum\limits_{i=1}^2\frac{\exp\left(-2r_{2i}\right)}{\sigma^2_{N_i}}\right)\exp\left(-\frac{2}{n}I\left(X^n;f^{(n)}_{E,1}(Y^n_1),f^{(n)}_{E,2}(Y^n_2)\right)\right)+\sigma^2_M\right).\label{19}
\end{eqnarray*}
Substitute (\ref{19}) into (\ref{13}) and then apply (\ref{9}),
\begin{eqnarray*}
&&I\left(f^{(n)}_{E,1}(Y^n_1);f^{(n)}_{E,2}(Y^n_2)\right)\\
&\geq&\frac{n}{2}\log\left(\frac{d_X+\sigma^2_M}{(d_1+\sigma^2_M)(d_2+\sigma^2_M)}\right)
\\&&+\frac{n}{2}\log\left(\sigma^2_XD_{3,\min}\left(\sum\limits_{i=1}^2\frac{\exp\left(-2r_{2i}\right)}{\sigma^2_{N_i}}\right)\exp\left(-\frac{2}{n}I\left(X^n;f^{(n)}_{E,1}(Y^n_1),f^{(n)}_{E,2}(Y^n_2)\right)\right)+\sigma^2_M\right)\\
&\geq&\frac{n}{2}\log\left(\frac{d_X+\sigma^2_M}{(d_1+\sigma^2_M)(d_2+\sigma^2_M)}\right)
\\&&+\frac{n}{2}\log\left(\sigma^2_XD_{3,\min}\left(\sum\limits_{i=1}^2\frac{\exp\left(-2r_{2i}\right)}{\sigma^2_{N_i}}\right)\exp\left(\frac{2}{n}I\left(f^{(n)}_{E,1}(Y^n_1);f^{(n)}_{E,2}(Y^n_2)\right)\right)\exp\left(-2(r_{11}+r_{21})\right)+\sigma^2_M\right),
\end{eqnarray*}
which can be rewritten as
\begin{eqnarray}
&&\exp\left(\frac{2}{n}I\left(f^{(n)}_{E,1}(Y^n_1);f^{(n)}_{E,2}(Y^n_2)\right)\right)\nonumber\\
&\geq&\frac{\left(d_X+\sigma^2_M\right)\sigma^2_M}{\left(d_1+\sigma^2_M\right)\left(d_2+\sigma^2_M\right)-\sigma^2_XD_{3,\min}\left(d_X+\sigma^2_M\right)\exp\left(-2\left(r_{11}+r_{21}\right)\right)\left(\sum\limits_{i=1}^2\frac{\exp\left(-2r_{2i}\right)}{\sigma^2_{N_i}}\right)}.\label{22}
\end{eqnarray}
Combining (\ref{23}) and (\ref{22}) yields that
\begin{eqnarray}
&&\exp\left(2\left(r_{11}+r_{21}\right)\right)\nonumber\\
&\geq&\frac{\sigma^2_X\sigma^2_M\left(\sigma^2_X+\sigma^2_M\right)}{D_3\left(\left(d_1+\sigma^2_M\right)\left(d_2+\sigma^2_M\right)-\sigma^2_XD_{3,\min}\left(d_X+\sigma^2_M\right)\exp\left(-2\left(r_{11}+r_{21}\right)\right)\left(\sum\limits_{i=1}^2\frac{\exp\left(-2r_{2i}\right)}{\sigma^2_{N_i}}\right)\right)},\nonumber
\end{eqnarray}
which can be further written as
\begin{eqnarray*}
r_{11}+r_{21}\geq\frac{1}{2}\log\frac{\sigma^2_X}{D_3}+\eta(\sigma^2_M,d_1,d_2,\zeta),
\end{eqnarray*}
where
\begin{eqnarray*}
&&\zeta=D_{3,\min}
D_3\left(\sum\limits_{i=1}^2\frac{\exp\left(-2r_{2i}\right)}{\sigma^2_{N_i}}\right),\\
&&\eta(\sigma^2_M,d_1,d_2,\zeta)=\frac{1}{2}\log\frac{\left(d_X+\sigma^2_M\right)\left(\zeta+\sigma^2_M\right)}{\left(d_1+\sigma^2_M\right)\left(d_2+\sigma^2_M\right)}.
\end{eqnarray*}
Calculus shows that
\begin{eqnarray*}
&&\hspace{-6.0in}\sup\limits_{\sigma^2_M}\eta(\sigma^2_M,d_1,d_2,\zeta)\\=\left\{\begin{array}{ll}
  \eta(\infty,d_1,d_2,\zeta)=0, \hspace{2.6in} \zeta\leq d_1+d_2-d_X\\
  \eta(0,d_1,d_2,\zeta)=\frac{1}{2}\log\frac{d_X \zeta}{d_1d_2}, \hspace{2.13in}\zeta\geq \left(\frac{1}{d_1}+\frac{1}{d_2}-\frac{1}{d_X}\right)^{-1}\\
  \eta(\hat\sigma^2_M,d_1,d_2,\zeta)=\frac{1}{2}\log\frac{(d_X-\zeta)^2}
{(d_X-\zeta)^2-[\sqrt{(d_X-d_1)(d_X-d_2)}-\sqrt{(d_1-\zeta)(d_2-\zeta)}]^2},  \mbox{ otherwise} \\
\end{array}\right.&\\
\end{eqnarray*}
where
\begin{equation*}
\hat\sigma^2_M=\frac{d_1d_2-d_X\zeta+\sqrt{(d_X-d_1)(d_X-d_2)(d_1-\zeta)(d_2-\zeta)}}{d_X+\zeta-d_1-d_2}.
\end{equation*}
\end{proof}

\section{Proof of Lemma 1}
Define the functions $y_1=\phi_1(x_1)$ and $y_2=\phi_2(x_2)$ via
the following parametric forms:
\begin{eqnarray*}
x_1&=&\frac{1}{2}\log\left(\frac{1}{\sigma^2_X}+\frac{1-\exp(-2\alpha_1)}{\sigma^2_N}\right)+\frac{1}{2}\log\sigma^2_X+\alpha_1, \\
y_1&=&\frac{1}{2}\log\left(\frac{1}{\sigma^2_X}+\frac{2-2\exp(-2\alpha_1)}{\sigma^2_N}\right)-\frac{1}{2}\log\left(\frac{1}{\sigma^2_X}+\frac{1-\exp(-2\alpha_1)}{\sigma^2_N}\right)+\alpha_1,
\end{eqnarray*}
and
\begin{eqnarray*}
x_2&=&
\frac{1}{2}\log\left(\frac{1}{\sigma^2_X}+\frac{2-2\exp(-2\alpha_2)}{\sigma^2_N}\right)-\frac{1}{2}\log\left(\frac{1}{\sigma^2_X}+\frac{1-\exp(-2\alpha_2)}{\sigma^2_N}\right)+\theta_2,\\
y_2&=&
\frac{1}{2}\log\left(\frac{1}{\sigma^2_X}+\frac{1-\exp(-2\alpha_2)}{\sigma^2_N}\right)+\frac{1}{2}\log\sigma^2_X+\alpha_2,
\end{eqnarray*}
where $\alpha_1,\alpha_2\geq 0$. Define
$\Omega=\{(R_1,R_2)\in\mathcal{R}^2_+:\phi_1(R_1)\leq
R_2\leq\phi_2(R_1)\}$. It is easy to check that $(R,R)$ is an
interior point of $\Omega$ for any $R>0$.

Let $\Gamma$ denote the line segment from $(R'_1,R'_2)$ to
$(R''_1,R''_2)$. It is clear that $D^*_3(R_1,R_2)$ must be a
convex function of $(R_1,R_2)$. Hence we have $\lambda_1
D^*_3(R'_1,R'_2)+\lambda_2 D^*_3(R''_1,R''_2)\geq D^*_3(R,R)$. If
the equality is achieved, then it implies that $D^*_3(R_1,R_2)$ is
linear on $\Gamma$.

It was computed in \cite{ChenCEO}\cite{SuccessiveCEO} that for any
$(R_1,R_2)\in\Omega$,
\begin{equation*}
R_1+R_2=\frac{1}{2}\log\left(\frac{\sigma^2_X}{D^*_3(R_1,R_2)}\left(\frac{2D^*_3(R_1,R_3)\sigma^2_X}{2D^*_3(R_1,R_3)\sigma^2_X-\sigma^2_X\sigma^2_N+D^*_3(R_1,R_2)\sigma^2_N}\right)^2\right),
\end{equation*}
or equivalently,
\begin{eqnarray*}
D^*_3(R_1,R_2)&=&\frac{2\sigma^4_X\sigma^2_N+\sigma^2_X\sigma^4_N+2\sigma^6_X\exp(-2(R_1+R_2))}{(2\sigma^2_X+\sigma^2_N)^2}\\
&&+\frac{2\sigma^4_X\exp(-(R_1+R_2))\sqrt{\sigma^4_X\exp(-2(R_1+R_2))+2\sigma^2_X\sigma^2_N+\sigma^4_N}}{(2\sigma^2_X+\sigma^2_N)^2},
\end{eqnarray*}
which is strictly convex with respect to $R_1+R_2$. Hence if
$R'_1+R'_2\neq R''_1+R''_1$, then $D^*_3(R_1,R_2)$ is strictly
convex on $\Omega\cap\Gamma$ and we must have $\lambda_1
D^*_3(R'_1,R'_2)+\lambda_2 D^*_3(R''_1,R''_2)> D^*_3(R,R)$. Note:
Since $(R,R)$ is an interior point of $\Omega$, $\Omega\cap\Gamma$
is not empty.

\section{Extreme Case}

\begin{eqnarray}
nR_1&\geq&H\left(f^{(n)}_{E,1}(Y^n_1)\right)=I\left(Y^n_1;f^{(n)}_{E,1}(Y^n_1)\right)=I\left(X^n,Y^n_1;f^{(n)}_{E,1}(Y^n_1)\right)\nonumber\\
&=&I\left(X^n,Y^n_1;f^{(n)}_{E,1}(Y^n_1),Y^n_2\right)-I\left(X^n,Y^n_1;Y^n_2\left|f^{(n)}_{E,1}(Y^n_1)\right.\right)\nonumber\\
&=&I\left(X^n;f^{(n)}_{E,1}(Y^n_1),Y^n_2\right)+I\left(\left.Y^n_1;f^{(n)}_{E,1}(Y^n_1),Y^n_2\right|X^n\right)\nonumber\\
&&-I\left(X^n,Y^n_1,f^{(n)}_{E,1}(Y^n_1);Y^n_2\right)+I\left(f^{(n)}_{E,1}(Y^n_1);Y^n_2\right)\nonumber\\
&=&I\left(X^n;f^{(n)}_{E,1}(Y^n_1),Y^n_2\right)+I\left(\left.Y^n_1;f^{(n)}_{E,1}(Y^n_1)\right|X^n\right)+I\left(Y^n_1;Y^n_2\left|X^n,f^{(n)}_{E,1}(Y^n_1)\right.\right)\nonumber\\
&&-I\left(X^n,Y^n_1;Y^n_2\right)+I\left(f^{(n)}_{E,1}(Y^n_1);Y^n_2\right).\label{99}
\end{eqnarray}

Now we bound each term separately. By data processing inequality,
we have
\begin{equation}
I\left(X^n;f^{(n)}_{E,1}(Y^n_1),Y^n_2\right)\geq I\left(X^n;\hat
X^n_3\right)\geq\frac{n}{2}\log\frac{\sigma_X^2}{D_3}.\label{100}
\end{equation}
Applying Lemma \ref{lemmaApp2} with $f^{(n)}_{E,2}(Y^n_2)=Y^n_2$,
we get
\begin{equation}
\frac{1}{\sigma_X^2}+\frac{1}{\sigma_{N_2}^2}+\frac{1-\exp\left(-\frac{2}{n}I\left(\left.Y^n_1;f^{(n)}_{E,1}(Y^n_1)\right|X^n\right)\right)}{\sigma_{N_1}^2}
\geq\frac{1}{\sigma_X^2}\exp\left(\frac{2}{n}I\left(X^n;f^{(n)}_{E,1}(Y^n_1),Y^n_2\right)\right).\label{101}
\end{equation}
Combining (\ref{100}) and (\ref{101}) and after simple
calculation, we obtain
\begin{equation}
I\left(\left.Y^n_1;f^{(n)}_{E,1}(Y^n_1)\right|X^n\right)\geq\frac{n}{2}\log\frac{\sigma_X^2\sigma_{N_2}^2D_3}{\left(D_3\sigma_X^2\sigma_{N_1}^2+D_3\sigma_X^2\sigma_{N_2}^2+D_3\sigma_{N_1}^2\sigma_{N_2}^2-\sigma_X^2\sigma_{N_1}^2\sigma_{N_2}^2\right)}
\end{equation}
Since $Y^n_1\rightarrow X^n\rightarrow Y^n_2$, it follows that
\begin{equation}
I\left(Y^n_1;Y^n_2\left|X^n,f^{(n)}_{E,1}(Y^n_1)\right.\right)=0
\end{equation}
and
\begin{equation}
I\left(X^n,Y^n_1;Y^n_2\right)=I\left(X^n;Y^n_2\right)=\frac{n}{2}\log\frac{\sigma_X^2+\sigma^2_{N_2}}{\sigma^2_{N_2}}.
\end{equation}
For the term $I\left(f^{(n)}_{E,1};Y^n_2\right)$, since
\begin{eqnarray*}
\frac{1}{n}\sum\limits_{t=1}^n{E(Y_2(t)-\hat
X_1(t))^2}&=&\frac{1}{n}\sum\limits_{t=1}^n{E(X(t)-\hat
X_1(t))^2}+\frac{1}{n}\sum\limits_{t=1}^n{EN_2^2(t)}\\&\leq&
D_1+\sigma^2_{N_2},
\end{eqnarray*}
by data processing inequality and then rate distortion theory, we
have
\begin{eqnarray}
I\left(f^{(n)}_{E,1}(Y^n_1);Y^n_2\right)\geq I\left(\hat
X^n_1;Y^n_2\right)\geq\frac{n}{2}\log\frac{\sigma_X^2+\sigma^2_{N_2}}{D_1+\sigma^2_{N_2}}.\label{102}
\end{eqnarray}
Now substituting (\ref{100})-(\ref{102}) back to (\ref{99}), we
get
\begin{equation}
R_1\geq\frac{1}{2}\log\frac{\sigma_X^4\sigma_{N_2}^4}{\left(\sigma_{N_2}^2+D_1\right)\left(D_3\sigma_X^2\sigma_{N_1}^2+D_3\sigma_X^2\sigma_{N_2}^2+D_3\sigma_{N_1}^2\sigma_{N_2}^2-\sigma_X^2\sigma_{N_1}^2\sigma_{N_2}^2\right)}.
\end{equation}

The main technical difference between the derivation here and the
one we used to prove the outer bound in Appendix II is the way to
lower bound
$I\left(f^{(n)}_{E,1}(Y^n_1);f^{(n)}_{E,2}(Y^n_2)\right)$. Since
for the extreme case it reduces to the problem of lower bounding
$I\left(f^{(n)}_{E,1}(Y^n_1);Y^n_2\right)$, we adopt a
straightforward approach as shown above, rather than the method of
Ozarow
\cite{Ozarow}.

\end{document}